\documentclass[12pt]{iopart}

\usepackage{iopams}
\expandafter\let\csname equation*\endcsname\relax

\expandafter\let\csname endequation*\endcsname\relax
\usepackage{physics}
\usepackage[x11names]{xcolor}
\usepackage{graphicx}
\usepackage{color}
\usepackage{hyperref}
\usepackage{cite}
\usepackage[normalem]{ulem}
\usepackage{comment}

\newcommand{\rcom}{\vb{R}}
\newcommand{\Jv}{\vb{J}}
\newcommand{\Lv}{\vb{\Lambda}}
\newcommand{\Ju}{\vu{J}}
\newcommand{\Sv}{\vb S}
\newcommand{\A}{S_2}
\newcommand{\qexp}{\varsigma}
\newcommand{\xv}{\vb{x}}
\newcommand{\yv}{\vb{y}}
\newcommand{\qv}{{\vb{q}}}
\newcommand{\kv}{{\vb{k}}}
\newcommand{\df}{\varphi}
\newcommand{\mob}{\mathcal{M}}
\newcommand{\qcross}{q_\ast}
\newcommand{\qh}{\hat{q}}
\newcommand{\qhcross}{\hat{q}_\ast}

\begin{document}
\title[Hyperuniformity in phase ordering]{Hyperuniformity in phase ordering: the roles of activity, noise, and non-constant mobility}

\author{Filippo De Luca$^1$, Xiao Ma$^1$, Cesare Nardini$^{2,3}$, Michael E. Cates$^1$}

\address{$^1$DAMTP, Centre for Mathematical Sciences, University of Cambridge, Wilberforce Road, Cambridge CB3 0WA, United Kingdom}
\address{$^2$Service de Physique de l'\'Etat Condens\'e, CEA, CNRS Universit\'e Paris-Saclay, CEA-Saclay, 91191 Gif-sur-Yvette, France}
\address{$^3$Sorbonne Universit\'e, CNRS, Laboratoire de Physique Th\'eorique de la Mati\`ere Condens\'ee, 75005 Paris, France}
\ead{fd408@cam.ac.uk}
\vspace{10pt}
\begin{indented}
\item[]5 July 2024
\end{indented}
\begin{abstract}
Hyperuniformity emerges generically in the coarsening regime of phase-separating fluids. Numerical studies of active and passive systems have shown that the structure factor $S(q)$ behaves as $q^\qexp$ for $q\to 0$, with hyperuniformity exponent $\qexp = 4$. 
For passive systems, this result was explained in 1991 by a qualitative scaling analysis of Tomita, exploiting isotropy at scales much larger than the coarsening length.
Here we reconsider and extend Tomita's argument to address cases of active phase separation and of non-constant mobility, again finding $\qexp=4$.
We further show that dynamical noise of variance $D$ creates a transient $\qexp = 2$ regime for $\hat q\ll \qhcross \sim \sqrt{D} t^{[1-(d+2)\nu]/2}$, crossing over to $\qexp = 4$ at larger $\qh$. Here, $\nu$ is the coarsening exponent for the domain size $\ell$, such that $\ell(t)\sim t^\nu$, and $\qh \propto q \ell$ is the rescaled wavenumber. In diffusive coarsening $\nu=1/3$, so the rescaled crossover wavevector $\qhcross$ vanishes at large times when $d\geq 2$. The slowness of this decay suggests a natural explanation for experiments that observe a long-lived $\qexp = 2$ scaling in phase-separating {\em active} fluids (where noise is typically large). Conversely, in $d=1$,  we demonstrate that with noise the $\qexp = 2$ regime survives as $t\to\infty$, with $\qhcross\sim D^{5/6}$. (The structure factor is not then determined by the zero-temperature fixed point.) We confirm our analytical predictions by numerical simulations of continuum theories for active and passive phase separation in the deterministic case and of Model B for the stochastic case. We also compare them with related findings for a system near an absorbing-state transition rather than undergoing phase separation. A central role is played throughout by the presence or absence of a conservation law for the centre of mass position $\rcom$ of the order parameter field.
\end{abstract}

%
\vspace{2pc}

\vspace{1em}
\submitto{\JPCM}
%
\maketitle
%
%

\section{Introduction}
Hyperuniformity is a form of hidden order at large scales, which involves the anomalous suppression of density fluctuations \cite{Torquato2018}. In a system with a single scalar field $\phi(\xv)$, the degree of hyperuniformity can be quantified by considering the long-distance behaviour of the autocorrelation function, defined as:
\begin{align}
\label{eq:g}
C(\xv) &= \frac{1}{L^{d}}\int\df(\xv+\yv)\df(\yv)\dd[d]{\bf y},
\end{align}
or, equivalently, of the structure factor:
\begin{equation}
\label{eq:S}
    S(\qv) = \int\dd[d]{\xv}C(\xv)e^{i\qv\vdot\xv}\,.
\end{equation}
Here $\df(\xv)=\phi(\xv)-\bar\phi$
denotes the deviation of the field from its spatial average $\bar\phi=(1/L^d) \int \dd[d]{\bf x}\phi({\bf x})$, and $L$ is the linear dimension of the system.

We will primarily be interested in the behaviour of the angular average of $S$ which, with a slightly abuse of notation, we will denote as $S(q)$, with $q=|{\qv}|$. (When isotropy exactly holds, $S(\qv) = S(q)$, and $C(\xv) = C(r)$ with $r=|\xv|$.) In a hyperuniform system, the structure factor vanishes in the long-wavelength limit $q \to 0$. The strength of this suppression of density fluctuations is captured by the exponent $\qexp>0$, which characterises the small-$q$ behaviour as $S(q)\sim q^\qexp$.

In phase-separating systems, the coarsening dynamics at late times (known as phase ordering) shows scaling behaviour, meaning that the time evolution of the autocorrelation function in \eqref{eq:g} obeys
$C(r, t) = c\left({r}/\ell(t)\right)$,
where $\ell(t)$ is the typical length scale of the phase-separated domains, and $c(r/\ell)$ is a scaling function~\cite{Bray1994}. The scaling behaviour of the auto-correlation translates into a scaling of the structure factor:
\begin{equation}
\label{eq:S_scaling}
	S(q,t) = \ell(t)^d s(\qh),
\end{equation}
where $\qh = q/q_\ell$ is the dimensionless wavenumber, and $q_\ell = 2\pi/\ell$. We will refer to the scaling function $s(\qh)$ as the \emph{rescaled structure factor} in the following. As time proceeds, the domains of opposing phases grow in size (see, {\em e.g.}, Fig.~\ref{fig:mu}a below). This results in a power-law growth of the characteristic length scale $\ell$, with $\ell(t)\sim t^\nu$. The exponent $\nu$ is well-known to be $\nu=1/3$ for passive scalar models undergoing diffusive dynamics in the absence of any coupling to a momentum-conserving fluid motion~\cite{LS1961,Wagner1961,Bray1994}.

Hyperuniformity under phase ordering has been reported in a number of scalar field theories, including a passive Cahn-Hilliard equation \cite{Ma2017} and, more recently, field theories describing phase separation in active systems \cite{Zheng2023}. Both these numerical studies found that the structure factor vanishes with an exponent of $\qexp=4$ at small wavevectors. While several theoretical explanations for the quartic scaling of the structure factor have been provided for Cahn-Hilliard equations~\cite{Yeung1988,Furukawa1989,Tomita1991,Bray1994}, these often rely on the dynamics being passive (driven by a chemical potential) and address only the case of constant mobility. It is so far unclear why a quadratic decay of the structure factor, $S(q)\sim q^2$ (which would be compatible with analyticity at $q=0$, a property that is valid in absence of long-range correlations), is generically absent even in cases beyond those in which those two assumptions hold: that is, for models with either activity (whereby time-reversal symmetry is absent even when noise terms are included), or non-constant mobility, or both. The effect of dynamical noise at continuum level has in contrast been addressed in the passive case~\cite{Tomita1991,Furukawa1989}, concluding that phase ordering is governed in $d\geq 2$ by a low-temperature fixed point at which noise is irrelevant~\cite{Bray1994}. We will see that despite this, the noise cannot be ignored altogether when examining the low-$q$ limit of $S(q)$, because it breaks a conservation law (Section \ref{sec:noise}); this confirms and extends a result in \cite{Furukawa1989}  for the passive, constant-mobility case.

Concerning noise terms, we make a brief remark here about nomenclature of models, which can be quite confusing. The Cahn-Hilliard equation~\cite{Cahn1958} does not include noise; its noisy version was first studied by Cook~\cite{Cook1970} and is sometimes called the Cahn-Hilliard-Cook equation. Moreover, Cahn and Hilliard did not assume a quartic polynomial for the local part of the free energy (denoted $f(\phi)$ below)~\cite{Cahn1958}; when a specific form was required, they used a lattice gas model instead. To this they added a square-gradient free-energy correction. Thus a noiseless, diffusive, square-gradient model with any choice of $f(\phi)$ is governed by {\em a} Cahn-Hilliard equation. Choosing a quartic $f(\phi)$, and {\em also adding noise} (in the form set by detailed balance) gives Model B~\cite{Hohenberg1977}. This is {\em not} a Cahn-Hilliard equation, but a statistical field theory (or stochastic PDE). However, its noiseless limit is itself often used in phase ordering studies. The resulting noiseless Model B {\em is} a Cahn-Hilliard equation and strangely (given that it is not the case addressed in~\cite{Cahn1958}), it is often called {\em the} Cahn-Hilliard equation, particularly in the mathematical literature. Accordingly, we will ourselves use this nomenclature for noiseless Model B below.

In this work, we build on an article by Tomita~\cite{Tomita1991} that provides a deceptively simple explanation for the $s(\qh)\sim \qh^4$ scaling at small $\qh$ in noiseless Model B. This uses the idea that the system is isotropic at scales much larger than the coarsening scale $\ell$. We review this argument, carefully examining its assumptions (including unstated ones, such as the hyperuniformity of the initial state) and relaxing some of those assumptions substantially. This allows us to extend the argument to a much more general class of scalar phase ordering models. While some of our results were already known for the passive case, we place these here within a unified framework that directly generalises to active theories, to non-constant mobilities, and to noisy systems. This enables us to provide a theoretical foundation for the exponent $\qexp=4$ observed in a variety of deterministic models with coarsening dynamics, including but not limited to Refs.~\cite{Ma2017,Zheng2023}. Our results also create a solid foundation for future study of hyperuniformity in phase-ordering models, beyond the case of a single scalar field addressed here.

For the noisy case, we demonstrate that standard diffusive noise leads to the emergence of a rescaled crossover wavenumber $\qhcross$ between regimes with $\qexp = 2$ (at small $\qh$) and $\qexp = 4$. For $d\geq2$, and with $\nu=1/3$, this crossover value goes to zero with time according to $\qhcross\sim t^{[1-\nu(d+2)]/2}$. In $d = 1$ though, the $\qexp = 2$ scaling of the long-wavelength regime survives in the scaling limit. We also show that if the noise variance is suppressed from the usual diffusive noise by a further factor $q^2$ (giving so-called `Laplacian noise'), the $\qexp = 2$ regime disappears in all dimensions. To illustrate how our arguments apply to other theories beyond phase separation, we also consider a reaction-diffusion model whose continuum description away from criticality (or in $d>4$) can be exactly solved \cite{Ma2023}. While not identical, we find strong mechanistic similarities between our conclusions for phase separation and those for that model, especially in relation to the role of a conservation law for the `centre of mass' of the order parameter field.

The paper is structured as follows. In Section \ref{sec:det}, we discuss hyperuniform scaling in noise-free theories of phase ordering. We extend the argument of \cite{Tomita1991} to take into account non-constant mobilities and activity, highlighting the mechanisms involved in the suppression of the quadratic part of the structure factor and the selection of a hyperuniformity exponent of $\qexp = 4$. In Section \ref{sec:noise}, we turn to the effect of noise in the dynamics. We demonstrate that standard diffusive noise leads to a transient $\varsigma=2$ scaling in $d\geq 2$, which becomes persistent in $d=1$. We also discuss how our results are relevant beyond coarsening systems, considering the example of a reaction-diffusion model. We further discuss our results, and give an outlook, in Section \ref{sec:discussion}.

\section{Hyperuniformity in deterministic models of phase ordering}
\label{sec:det}

Here we discuss hyperuniform scaling in deterministic models of a coarsening scalar field, where noise enters only through the initial condition. In Sec.~\ref{sec:models-PS}, we set up the models of phase ordering studied in the rest of the paper, defining the notation needed. In Sec.~\ref{sec:IC}, we discuss the scaling of the structure factor within the initial state. In Sec.~\ref{sec:det-stress} we re-work Tomita's argument, originally developed in~\cite{Tomita1991} for the Cahn-Hilliard equation with constant mobility, and generalise it to all cases in which the density flux can be written as the divergence of a local tensor, assuming (as Tomita implicitly did) hyperuniform initial conditions. We then extend Tomita's argument to general initial conditions in~\ref{sec:det-stress2}. In Sec.~\ref{sec:deterministic-mob} we discuss the case of non-constant mobilities in both passive and active models. In Sec. \ref{sec:2-num} we test all the predictions via numerical simulations. In Sec. \ref{sec:CoM} we briefly discuss the role played by an exact or approximate conservation law for the centre of mass of the order parameter field.

\subsection{Continuum models of phase-separating systems}\label{sec:models-PS}
We consider systems described by a single, conserved, scalar order parameter $\phi(\xv)$. We take the dynamics of $\phi$ to be purely diffusive, without coupling to a conserved momentum~\cite{Bray1994}. Its time evolution is then governed by a continuity equation:
\begin{equation}
\label{eq:continuity}
    \partial_t\phi = -\div\Jv + \div\Lv\,,
\end{equation}
with flux $\Jv(\xv,t)$ and $\Lv$ a zero-mean Gaussian random vector of covariance
\begin{equation}
\label{eq:Lambda_var}
\expval{\Lambda_i(\xv,t)\Lambda_j(\yv,t^\prime)} = 2D\mob(\phi)\delta_{ij}\delta(\xv-\yv)\delta(t-t^\prime),
\end{equation} 
with an everywhere-positive mobility $\mob(\phi)$, and $D$ a `noise temperature' that may or may not be thermal in origin~\cite{Cates2018}. Here and elsewhere, we use indices to refer to the components of a bold-face vector.
We first focus on the deterministic case, $D=0$,  postponing the analysis of dynamical noise  effects to Sec.~\ref{sec:noise}. 

In the passive case, the flux is given in terms of the mobility and the gradient of a chemical potential $\mu$, which itself is the functional derivative of a free energy functional $\mathcal{F}$, as:
\begin{equation}
\label{eq:Jpass}
\Jv_\mathrm{pass} = -\mob(\phi)\grad\mu;\qquad
\mu= \fdv{\mathcal{F}}{\phi}\,.
\end{equation}
For simplicity, we choose a free energy $\mathcal{F}$ of square-gradient $\phi^4$ form,
\begin{equation}\label{eq:free-energy}
\mathcal{F} = \int\dd[d]{\xv}\left(f(\phi)+\frac{K}{2}(\grad\phi)^2\right);\qquad f(\phi)=-\frac{a}{2}\phi^2+\frac{b}{4}\phi^4,
\end{equation}
but our results can be easily extended to more general $\mathcal{F}$. Here, $a$, $b$ and the stiffness $K$ are positive parameters of the theory. Note that, given \eqref{eq:free-energy}, $\phi$ denotes the deviation of the compositional order parameter $\phi$ from its value at the (mean-field) critical point where (without noise) the two separating phases  first become identical. For brevity, we will sometimes refer to $\phi$ as just `the density' in the following.

The resulting passive dynamics for $\phi$ (with $\Jv=\Jv_\mathrm{pass}$) obeys the Cahn-Hilliard equation, which is the noiseless limit of Model B \cite{Cahn1958,Cahn1965,Hohenberg1977,Bray1994,Cates2018}. For $a>0$, $\mathcal{F}$ is minimised by phase separation for global mean densities $\bar\phi$ obeying $|\bar\phi|<\phi_B=(a/b)^{1/2}$. Here $\pm\phi_B$ are the binodal densities. However, a near-homogeneous initial state is metastable for $\phi_B>|\bar\phi|>\phi_S = \phi_B/\sqrt{3}$ where $\pm\phi_S$ are the spinodal densities; phase separation cannot then be reached from a homogeneous starting state with small initial noise. In contrast, for $|\bar\phi|<\phi_S$ the homogeneous state is locally unstable, so infinitesimal initial noise leads to phase separation under deterministic Cahn-Hilliard evolution~\cite{ChaikinLubensky1995,Onuki2002,Cates2018}.

In active systems, detailed balance is broken at the scale of the demixing species, implying that the flux cannot in general be derived from a free energy functional as in \eqref{eq:free-energy}. 
This can be addressed by adding to Model B low-order terms (in a Landau-Ginzburg sense) that break time-reversal symmetry~\cite{Wittkowski2014,nardini2017entropy}. The corresponding minimal model, known as Active Model B+ (AMB+) \cite{Tjhung2018}, contains all terms that independently break detailed balance to order ${O}(\nabla^3\phi^2)$. Without noise, the additional active flux reads:
\begin{equation}
\label{eq:Jact}
\Jv_\mathrm{act} = -\mob(\phi)\left\{\lambda\grad[(\grad\phi)^2] - \zeta(\laplacian\phi)\grad\phi\right\}.
\end{equation}
Here, the parameters $\lambda$ and $\zeta$ control the strength of activity. AMB+ was studied so far in the case of constant mobility $\mob(\phi)=\mob$, analogously to the standard choice that is made for passive fluids. In the following, we will relax this assumption, allowing in (\ref{eq:Jpass},\ref{eq:Jact}) a generic density-dependent mobility. 
Note that the phenomenology of AMB+ is richer than passive Model B, because the presence of the $\zeta$ term can reverse the Ostwald process~\cite{Tjhung2018}, alter nucleation scenarios~\cite{cates2023classical}, and create new capillary instabilities~\cite{fausti2021capillary}. However we will not consider such regimes in this paper, limiting ourselves instead to cases where standard phase ordering is found. These cases include an earlier model, Active Model B (AMB) which has $\lambda \neq 0$, $\zeta = 0$~\cite{Wittkowski2014}.

We now decompose the flux into a part that can be written as the divergence of a local rank-two tensor $\Sv$, and a remainder $\Ju$:
\begin{equation}
\label{eq:J_decomposition}
\Jv=\Jv_\mathrm{pass}+\Jv_\mathrm{act} = -\div\Sv + \Ju.
\end{equation}
This is reminiscent of the decomposition of a force vector field into surface and body forces in continuum mechanics. Note however that $\Sv$ is not the mechanical stress tensor ${\bf \Sigma}$, which for a passive scalar obeys $\div{\bf \Sigma} = -\phi\nabla\mu$ \cite{Bray1994}. Moreover, the requirement that $\Sv(\xv)$ be local (\emph{i.e.}, it only depends on the field and its gradients evaluated at the point $\xv$) is crucial here. Indeed, by Helmholtz decomposition, 
any vector field $\Jv$ can be written (without a $\Ju$ term) as the divergence of {\em some} second-rank tensor $\tilde{\Sv}$. However,  $\tilde{\Sv}$ is then nonlocal (and found by Biot-Savart/Coulomb-type integrals). 

For an arbitrary (positive) local mobility function $\mob(\phi)$, $\Sv$ in \eqref{eq:J_decomposition} can now be written as:
\begin{equation}
\label{eq:Stot}
S_{ij} = \mob(\phi)\left[\mu + (\lambda-\zeta/2)(\grad\phi)^2 -\bar\mu\right]\delta_{ij} - \mob(\phi)\zeta[\partial_i\phi\partial_j\phi - (\grad\phi)^2\delta_{ij}],
\end{equation}
where $\bar\mu$ is a constant reference value (exploited later). In a passive system with constant mobility, modulo a shift by $\bar\mu$, the tensor $\Sv$ reduces to the chemical potential, $S_{ij}/{\mob} = (\mu-\bar\mu)\delta_{ij}$; similarly, for $\zeta = 0$ and $\lambda\neq 0$, it reduces to a non-equilibrium chemical potential, defined in~\cite{Wittkowski2014}, $S_{ij}/{\mob} = [\mu+\lambda(\grad\phi)^2-\bar\mu]\delta_{ij}$. For non-vanishing $\zeta$, however, $\Sv$ becomes non-diagonal. This reflects the absence in general of any {\em local} non-equilibrium chemical potential in AMB+ \cite{Tjhung2018}. 
The remaining current $\Ju$ in \eqref{eq:J_decomposition} reads:
\begin{align}
\Ju = \mob^\prime(\phi)\left[\mu + (\lambda-\zeta/2)(\grad\phi)^2-\bar\mu \right]\grad\phi,\label{eq:Ju}
\end{align}
where the prime ($^\prime$) denotes a $\phi$ derivative. Note that the decomposition of $\Jv$ into $\Sv$ and $\Ju$ is not unique: for example, the constant $\bar\mu$ can be chosen at will. Importantly, our choice of $\Ju$ vanishes for constant mobility, in both the passive and active models under study here. For now we focus on this case, deferring non-constant mobilities such that $\Ju\neq {\bf 0}$ to Section \ref{sec:deterministic-mob}.

\subsection{Initial condition}\label{sec:IC}
The initial state of the system is taken to be uniform up to weak noise, whose role is to trigger the linear spinodal instability. (Dynamical noise is absent.) For example, in the passive case we may choose the initial condition to be drawn from the Boltzmann distribution of  \eqref{eq:free-energy} with initial noise $D_0$, and `pre-quench' parameters $a = a_0 = -\bar a$ and $b=0$. Then the initial, pre-quench structure factor is $S(q) = D_0/(\bar a+Kq^2)$.
We define the Taylor expansion in $q$ of the subsequent time-evolved structure factor as
\begin{equation}\label{eq:pre-quench}
S(q,t) = S_0(t)+S_2(t)q^2+S_4(t)q^4 +O(q^6).
\end{equation}
In this expansion and throughout this work, we exclude odd (and non-integer) powers of $q=\abs{\qv}$; this analyticity assumption is tantamount to ruling out long-range correlations (for example, a term $\sim |q|^n$ in $S(q)$ with odd $n$ would correspond to a power-law tail in the autocorrelation $C(r)\sim r^{-(d+n)}$)~\cite{Torquato2018}.
Crucially, $S_0(t)$ is defined as $S(q\to 0^+,t)$, not as $S(0,t)$, since according to \eqref{eq:g} and \eqref{eq:S} the latter vanishes identically even in non-hyperuniform systems. We then have the initial conditions 
\begin{equation}\label{eq:ICs}
S_0(0) = D_0/\bar a,\quad S_2(0) = -D_0K/\bar a^2,\quad S_4(0) = D_0K^2/\bar a^3.
\end{equation}
Infinitesimal noise ($D_0\to 0^+$), is enough to trigger phase separation within the spinodal region. The initial exponential growth of fluctuations crosses over into the phase-ordering regime once sharp interfaces emerge between phases of saturated density $\pm\phi_B$.

In what follows, we do not need \eqref{eq:ICs} directly, only the fact that $S(q)$ is initially small (compared to the values later obtained during phase ordering) and that its Taylor expansion does not contain explicit dependences on the system size $L$, assumed large. This applies for any initial condition created by quenching from a uniform phase. The sense in which hyperuniformity then emerges during phase ordering is not that $S_0(t)$ tends strictly to zero but that it remains small, while the rest of the structure factor grows to much larger values. These large values stem from the eventual order-one (as opposed to initial order $D_0$) density variations between domains at density $\pm\phi_B$. The rescaled structure factor $s(\qh)$ then encodes the spatial arrangement of these domains. 

To formalise this statement, we consider the rescaled structure factor $s(\qh)$ of \eqref{eq:S_scaling}. Its Taylor expansion reads:
\begin{equation}
\label{eq:s_expansion}
    s(\qh) = s_0\qh^0 + s_2\qh^2 + s_4\qh^4 + O(\qh^6),
\end{equation}
where the expansion coefficients are related to those of $S(q)$ by:
\begin{align}
\label{eq:rescaled_S_coeff}
    s_n = (2\pi)^n\ell^{-(d+n)}S_n.
\end{align}
In the scaling regime for phase ordering, $s(\qh)$ is time-independent. Hence, for a $\qh^n$ part to be present in the scaling function, $s_n$ must be time-independent, which requires $S_n$ to grow in time at least as fast as $\ell^{d+n} \sim t^{(d+n)\nu}$. Now, in view of the (mass) conservation law on $\phi$, it follows that $S_0=\mathrm{const.}$ for finite times. (In particular, in the linear instability regime the growth rate vanishes at low $q$.) Thus, because of \eqref{eq:rescaled_S_coeff}, $s_0 \to 0$ in time as $\ell^{-d}$. Hyperuniformity therefore is expected; the question is, what is the exponent $\varsigma$? In what follows we will show that $S_2(t)$ also remains small, leading to $\varsigma = 4$. This result, at least for constant mobility, can be explained using a separate conservation law for the centre-of-mass vector $\rcom$ introduced below. (This conservation has been overlooked in much of the phase-ordering literature, though not by Tomita~\cite{Tomita1991}). We will first treat the case of hyperuniform initial conditions, where $S_0(0)=0$ exactly, but will discuss how the arguments extend to general $S_0(0)$ in Sec.~\ref{sec:det-stress2} below.

\subsection{Hyperuniform initial conditions and constant mobility}
\label{sec:det-stress}
In this Section, we closely follow Tomita~\cite{Tomita1991}, who implicitly considered only the case where $S_0(0) = 0$, {\em i.e.}, exactly hyperuniform initial conditions, and also assumed constant mobility, so that $\Ju = 0$ in \eqref{eq:J_decomposition}. 
We show that under these conditions $S_2(t)$ stays small. Unlike Tomita's, our analysis is not specific to the Cahn-Hilliard equation. The main idea is that the scaling of the structure factor at small $q$ follows from the isotropy of the system at sufficiently large scales. Our treatment here assumes no-flux boundary conditions in a fixed cubic domain. However, one can always suppose such boundary conditions to be applied arbitrarily far from a smaller cubic subdomain within which $S(q)$ is defined and studied. We presume similar results could be derived with periodic boundary conditions, and have done so in some cases, but do not pursue this aspect here. The results of this and the next Section explain the numerical findings of~\cite{Ma2017,Wilken2023}, observing that $S(q)\sim q^4$ in the phase ordering regime of the Cahn-Hilliard model, and also in Active Model B and B+ with constant $\mob$. All of these have $\Ju = \bf{0}$ by virtue of \eqref{eq:Ju}.

We start by expanding the exponential in \eqref{eq:S} to fourth order in $\qv$:
\begin{align}\label{eq:Ces-Sq}
S(\qv) &= -\frac{1}{2}\int\dd[d]{\xv}(\qv\vdot\xv)^2C(\xv) +\frac{1}{4!}\int\dd[d]{\xv}(\qv\vdot\xv)^4C(\xv) + {O}(q^6)\,,
\end{align}
where odd powers in $\qv$ vanish due to $C(\xv) = C(-\xv)$. We then integrate over angular coordinates, defining:
\begin{equation}\label{eq:Ces-Sq-1}
    S(q) = \frac{1}{A_{d-1}}\int\dd[d-1]{\Omega} S(\qv),
\end{equation}
where $A_{d-1} = 2\pi^{d/2}/\Gamma(d/2)$ is the surface area of the unit $(d-1)$-sphere, and $\dd[d-1]{\Omega}$ its integration measure.  From \eqref{eq:Ces-Sq} and \eqref{eq:Ces-Sq-1}, we have:
\begin{subequations}
\begin{align}
\label{eq:S_expansion}
S(q) &= S_2 q^2+S_4 q^4+ {O}(q^6),\\
S_2 &=-\frac{1}{2d\,L^d}\int\dd[d]{\xv}\int\dd[d]{\bf y}\df(\xv+\yv)\df(\yv)|\xv|^2,\label{eq:S_expansion-S2}\\
S_4 &=\frac{3}{4!d(d+2)\,L^d}\int\dd[d]{\xv}\int\dd[d]{\bf y}\df(\xv+\yv)\df(\yv)|\xv|^4\,.\label{eq:S_expansion-S4}
\end{align}
\end{subequations}
A similar expansion of the structure factor in moments of the correlation function can be done for particle systems as well \cite{Torquato2022}.

We now look for a relation between the coefficient $\A$ and the asymmetry of the density distribution, as encoded in the following `centre-of-mass' vector:
\begin{equation}
\label{eq:rcom_df}
    \rcom:=\frac{1}{L^d}\int\dd[d]{\xv}\xv\,\df(\xv).
\end{equation}
This vector can be seen as a dipole moment of the density fluctuations, pointing towards regions of higher density with magnitude set by the strength of the asymmetry. For a fully homogeneous system, $\df = 0$ everywhere, and $\rcom = \bf{0}$. In contrast, for a macroscopic phase-separated state, $|{\rcom}|$ scales as the linear dimension $L$ of the system.

Shifting the integral over $\xv$ in \eqref{eq:S_expansion-S2}, taking careful account of the limits of integration (see \ref{app:noflux}), gives the result:
\begin{align}
\A &= -\frac{1}{2d\,L^d}\int\dd[d]{\xv}\int\dd[d]{\bf y}(\xv-\yv)^2\df(\xv)\df(\yv)= \frac{1}{d\,L^d}\int\dd[d]{\xv}\df(\xv)\xv\vdot\int\dd[d]{\bf y}\yv\df(\yv)\,.\nonumber
\end{align}
Here we have used $\int \dd[d]{\bf x} \df({\bf x}) = 0$ to get the second equality. It follows that
\begin{align}
\A = \frac{L^d}{d}|\rcom|^2\,. \label{eq:A_R}
\end{align}
Note that this result of Tomita uses \eqref{eq:S_expansion} and so is exact only to the extent $S_0$ can be neglected: it {\em assumes} hyperuniformity, although Tomita does not mention this fact. We will discuss how this relation changes for $S_0\neq 0$ in Sec.~\ref{sec:det-stress2} below.

To establish that the hyperuniformity exponent is $\varsigma = 4$ and not $\varsigma  = 2$ for phase ordering, we now show that the quadratic coefficient $S_2(t)$ remains small under time evolution, because its time derivative vanishes as a negative power of $L$ for large system sizes $L$. We start from \begin{align}
\label{eq:A_dt}
    \partial_t \A &= \frac{L^d}{d}\partial_t|\rcom|^2 = \frac{2L^d}{d}\rcom\vdot\vb{I},
\end{align}
with $\partial_t\rcom=\vb{I}$ and $\vb{I}$ defined as
\begin{align}
\vb{I}  &:= -\frac{1}{L^d}\int_V\dd[d]{\xv}\xv\div\Jv(\xv) = \frac{1}{L^d}\int_V\dd[d]{\xv}\Jv(\xv)
\nonumber\\
&=
-\frac{1}{L^d}\int_{\partial V}\dd[d-1]{\xv}\Sv(\xv)\vdot\vu{n} + \frac{1}{L^d}\int_V\dd[d]{\xv}\Ju(\xv)\,.\label{eq:I}
\end{align}
Here we have used our boundary condition that $\Jv\vdot\vu{n}=0$ on the boundary $\partial V$ with $\vu{n}$ the outward surface normal there. Since also $\Ju={\bf 0}$, the volume term vanishes, so that the only contribution to $\vb{I}$ is a surface integral, whose integrand is a local function of $\phi$ and its derivatives. (Thus the integrand is $O(L^0)$.)

\begin{figure}
    \centering
    \includegraphics[width=\textwidth]{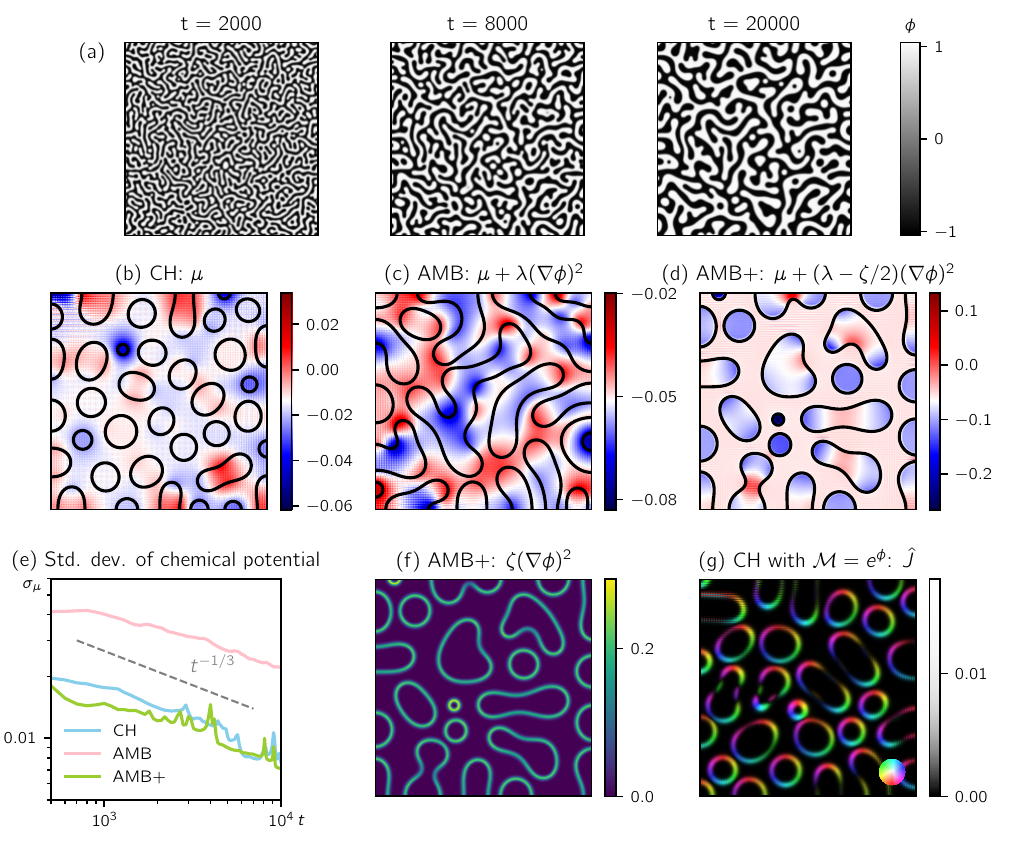}
    \caption{Coarsening dynamics of active and passive scalar field theories: Cahn-Hilliard (CH), AMB with $\lambda=-1$, $\zeta=0$ and AMB+ with $\lambda=0$, $\zeta=2$, all with $a=b=0.25$ and $K=1$, with $\mob=1$ in {(a-f)}, with $L=1024$ in (a) and $L=256$ in (b-g). A no-flux boundary condition is used in all cases. \textbf{(a)} Spinodal decomposition and coarsening in CH with $\bar\phi=0$; the other CH simulations here have $\bar\phi=0.2$. \textbf{(b-d)} The passive chemical potential $\mu=f^{\prime\prime}(\phi)-K\laplacian\phi$, and its generalisations for AMB(+), each acquire curvature-dependent variations of order $1/\ell$ to their constant averages. In \textbf{(d)}, $\zeta\neq 0$ leads to jumps at the interfaces of order $\ell^{-1}$. \textbf{(e)} According to their scaling, the standard deviations of the generalised chemical potentials in (b-d) from their spatial averages decay as $\ell^{-1}$, which for $\nu=1/3$ gives $t^{-1/3}$. \textbf{(f)} The isotropic part of the second term in \eqref{eq:Stot} is shown and, as expected, it is localised to the interfaces. \textbf{(g)} CH with exponential mobility. $\Ju$ is shown here, with colour indicating its direction (see colour wheel) and brightness its norm.} \label{fig:mu}
\end{figure}

We now consider the scaling of this surface integral, based on the properties of the second-rank tensor $\Sv(\xv)$ given in \eqref{eq:Stot}. The first term of that equation can be identified with the non-equilibrium chemical potential arising for flat interfaces in AMB+ \cite{Tjhung2018}. In the absence of curvature, this chemical potential is constant (and local). For $d\geq 2$, interfacial curvature leads to $O(\ell^{-1})$ corrections: in the passive case, these arise due to Laplace pressure, which changes the chemical potential by an amount $\sim\sigma/\ell$ where $\sigma$ is the interfacial tension. In the active case, this is still true, but with $\sigma$ now representing a non-equilibrium pseudotension~\cite{Tjhung2018}. For $\zeta\neq 0$, the non-equilibrium chemical potential additionally acquires a jump at any curved interface, which however is still of order $O(\ell^{-1})$~\cite{Tjhung2018}. In Fig.~\ref{fig:mu}b-d, we plot the generalised potential for all these cases. Figure \ref{fig:mu}e gives further numerical evidence for the scalings discussed here.

Note that any spatially uniform isotropic contribution to $S_{ij}$, denoted $\bar{S}\delta_{ij}$, will not contribute to the integral $\vb{I}$ because $\int_{\partial V}\vu{n}\dd[d-1]{\xv} = {\bf 0}$, so we can subtract the spatial average of the first term in \eqref{eq:Stot} from the integrand. (Inclusion there of the arbitrary reference value $\bar \mu$ makes this explicit.) We now appeal to large scale isotropy to assert that, until such time as $\ell(t)\simeq L$, the spatial fluctuations of the second-rank tensor $\Sv(\xv)$ around its average have no correlation with the outward normal vector on the boundary $\vu{n}$. In this case the surface integral is a sum over random, zero-mean contributions, each of amplitude $O(\ell^{-1})$, correlated patchwise on domains of area $\ell^{d-1}$. The resulting contribution in \eqref{eq:I} then scales as $|{\vb{I}}|\lesssim L^{-d}(L/\ell)^{(d-1)/2}\ell^{d-1}\ell^{-1} \sim L^{-(d+1)/2}\ell^{(d-3)/2}$. 

It is important that this estimate, like others involving the symbol $\lesssim$ that appear below, represents an upper bound (modulo dimensionless prefactors). Hence in this case, our argument says that $|{\vb{I}}|$ is bounded above by some quantity scaling as $L^{-(d+1)/2}\ell^{(d-3)/2}$ in the limit where $L\gg\ell\gg\xi$. (We suppress powers of the interface width $\xi$, using which the correct overall dimensions of any estimated quantity can be restored.) As shown below, this bound ensures $\varsigma = 4$ for the Cahn Hilliard equation~\cite{Tomita1991}. 

The second term in \eqref{eq:Stot} is an additional contribution arising in Active Model B+, localised at the interfaces between phases (see Fig.~\ref{fig:mu}f). Under Neumann boundary conditions ($\grad\phi\vdot\vu{n} = 0$), only the isotropic part of $S_{ij}$ contributes to $\vb{I}$. Up to corrections of relative order $\ell^{-1}$, and at scales larger than $\xi$, this term gives in \eqref{eq:I} a surface integrand whose magnitude is proportional to the contour length density with which the interface between phases intersects the system boundary, and whose direction is the normal $\vu{n}$ to that boundary. We can again subtract the average of this term in $\vb{I}$, since it cancels vectorially from the surface integral, which again leaves a sum of patchwise correlated random contributions: each patch now gives a contribution of order $\xi\ell^{d-2}$, where $\xi$ is the interfacial width and $\ell^{d-2}$ the contour length of the interface in the patch. This gives a total contribution of order $|{\vb{I}}|\lesssim L^{-d}(L/\ell)^{(d-1)/2}\ell^{d-2}\xi \sim L^{-(d+1)/2}\ell^{(d-3)/2}$. This is the same scaling as for the previous term.

For an initial $S_2\sim O(L^0)$, the relation \eqref{eq:A_R} gives $|{\rcom}|\sim L^{-d/2}$ at $t=0$. The dynamics leaves this scaling unaffected, since we have shown that $|\partial_t{\rcom}| = |{\vb{I}}|\sim L^{-(d+1)/2}\ll|{\rcom}|$. As discussed further in Section
\ref{sec:CoM} below, this effectively embodies a conservation law for $\rcom$. Likewise, $\partial_t S_2\lesssim L^{d} |\rcom|\cdot|\vb{I}| \lesssim L^{-1/2} \ll S_2$. Accordingly, $S_2$ stays close to its (order unity) initial value under the time evolution. Hence, because of \eqref{eq:rescaled_S_coeff}, $s_2$ decays as $s_2\sim t^{-\nu(d+2)}$, and is negligible at late times.

As emphasised by Tomita~\cite{Tomita1991}, there is no counterpart to the above arguments in the case of $S_4$, essentially because the time evolution of this quantity is not controlled by the divergence of any second-rank tensor. Thus, the quartic term in \eqref{eq:S_expansion} will grow to dominate the quadratic one in a system undergoing phase separation, and continue to dominate throughout the phase-ordering regime, until finally the domain scale $\ell(t)$ approaches the system size $L$.  During phase ordering, when $S(q)$ has become so large that its initial value can be neglected, the system will thus become effectively hyperuniform with exponent $\varsigma = 4$.

\subsection{The role of non-hyperuniform initial conditions}\label{sec:det-stress2}
In the previous section, we have shown that $S_2$ conserves its initial value in coarsening models with constant mobility, under the asssumption of hyperuniform initial conditions, so that $S_0 = 0$. Here, we relax this assumption, showing that even when $S_0 \neq 0$ the quadratic coefficient $S_2$ cannot keep up with the growth of the quartic term, thus resulting in a quartic scaling function $s(\qh)\sim \qh^4$ in the coarsening regime.

When $S_0$ is non-vanishing, the Taylor expansion in the form of \eqref{eq:S_expansion} is no longer valid. This form was explicitly used by Tomita~\cite{Tomita1991}, and in doing so he implicitly assumed hyperuniform initial conditions. (One might speculate that he overlooked the distinction, made above after \eqref{eq:pre-quench}, between $S_0\equiv S(q\to 0^+)$ and $S(0)$, of which only the latter vanishes identically.)
Its breakdown means that the relation \eqref{eq:A_R} between the centre of mass vector and the structure factor no longer holds. In \ref{app:S0}, we show how including $S_0$ modifies that relation, which becomes:
\begin{equation}
\label{eq:A_R_S0}
\rcom^2 \sim L^{2-d}S_0 + L^{-d}\A.
\end{equation}
Hence, for non-hyperuniform initial conditions with $S_0\sim O(D_0L^0)$, the scaling $|\rcom|\sim L^{-d/2}$ is changed to $|\rcom|\sim L^{1-d/2}S_0^{1/2}$ at $t=0$. The arguments for the scaling of $\vb{I}$ do not lose their validity, so that $\rcom$ is still effectively conserved in a large system: we have $|\partial_t\rcom| \sim L^{-(d+1)/2}$ so that still $|\partial_t\rcom| \ll |\rcom|$. Furthermore, as a consequence of the separate conservation law on the order parameter $\phi$, which suppresses all evolution for $q\to 0$, we can assume $\partial_tS_0 = 0$,  so that \eqref{eq:A_dt} continues to hold.

Using the scaling \eqref{eq:A_R_S0} for $\rcom$, the arguments given in the previous section suggest that $\partial_t\A \lesssim S_0^{1/2}L^{1/2}\ell^{(d-3)/2}$.  Integrating this relation and using \eqref{eq:rescaled_S_coeff} yields:
\begin{equation}
s_2 \lesssim S_0^{1/2}L^{1/2}\ell^{-(d+7)/2+1/\nu}. \label{eq:wrong}
\end{equation}
At first sight, this bound from $\rcom$-conservation now looks useless: it diverges as  $L\to \infty$ at fixed $\ell$, a limit in which $s_2$ is certainly {\em not} divergent. 
However the bound in \eqref{eq:wrong} can be refined into a useful one by a causality argument. This takes into account how a change in some neighbourhood does not affect distant parts of the system instantaneously, but instead requires propagation of information to those regions. Such propagation sets a causality length scale that grows in time as $L_\mathrm{eff}\sim t^\gamma \sim \ell^{\gamma/\nu}$. Since the behaviour in any neighbourhood cannot be influenced by events beyond the causality scale, $L_\mathrm{eff}$ can safely replace $L$ as an effective system size in bounds such as \eqref{eq:wrong}. Importantly, the original bound is not negated: we need make no such replacement in cases where the power of $L$ is helpfully negative (as arose for hyperuniform initial conditions in Sec.~\ref{sec:det-stress}).

For a system where information propagates at fixed speed $c$ ({\em e.g.}, via light or sound),  $L_\mathrm{eff} = ct$ and hence $\gamma=1$. However, for phase separation without a conserved momentum there are no under-damped modes and the fastest possible information transport is via diffusive scaling. This sets an upper bound of $\gamma\le1/2$ for all theories considered here. (Note that one might argue further that in phase ordering, information cannot move much faster than interfaces do, in which case $\gamma =\nu$. However, the weaker but more rigorous diffusive bound will suffice for our purposes.) For times within the scaling regime, one then has $\ell(t)\ll L_\mathrm{eff}(t)\ll L$ with the first two of these independent of system size  in the large $L$ limit.

Using $L_\mathrm{eff}$ as an effective system size instead of $L$ in \eqref{eq:wrong}, our improved bound is
\begin{equation}
s_2 \lesssim S_0^{1/2}\ell^{[(2+\gamma)/\nu-(d+7)]/2}.
\end{equation}
For $\nu = 1/3$ and $\gamma = 1/2$, this gives $s_2 \lesssim \ell^{(1-2d)/4}$, which falls to zero as $\ell$ grows with time, for any dimension $d$. Hence, we have shown that, even accounting for a non-hyperuniform initial condition $S_0\neq 0$, coarsening models with constant mobility lead to a scaling function $s(\qh)$ with quartic hyperuniform scaling, corresponding to $\varsigma = 4$.

\subsection{The role of density-dependent mobility}\label{sec:deterministic-mob}
We next turn to the case of $\phi$-dependent mobility $\mob(\phi)$, and start by decomposing the integral $\vb{I}$ as in \eqref{eq:I}. Choosing $\bar\mu$ as the constant value that the non-equilibrium chemical potential takes on for flat interfaces \cite{Tjhung2018}, we note that $\hat\mu := \mu+(\lambda-\zeta/2)(\grad\phi)^2-\bar\mu$, appearing in (\ref{eq:Stot},\ref{eq:Ju}) is of order $O(\ell^{-1})$, so that the arguments presented in Sec.~\ref{sec:det-stress} for the surface integral in \eqref{eq:I} retain their validity. On the other hand, with $\phi$-dependent mobility we have $\Ju\neq \vb{0}$ (see Fig.~\ref{fig:mu}g), so that $\vb{I}$ also contains a volume integral,
\begin{equation}
\label{eq:I_Ju2}
\vu{I} := \frac{1}{L^d}\int_V\dd[d]{\xv}\Ju(\xv) = \vu{I}_\mathrm{bulk} + \vu{I}_\mathrm{interfaces}\,,
\end{equation}
whose presence invalidates {\em a priori} the $\rcom$-conservation arguments of the previous two Sections.
We have in \eqref{eq:I_Ju2} split the integral into a contribution coming from the bulk regions comprising coarsening domains, and one from the interfaces between them. We now consider the scaling of these two terms.

In both bulk phases, the field $\phi$ differs from one or other binodal value by terms of order $\ell^{-1}$, with spatial variations on a length scale $\ell$, so that we can estimate $|\grad\phi| \sim \ell^{-2}$. On the other hand, the non-equilibrium chemical potential $\hat{\mu}$ is of order $\ell^{-1}$, so that the integrand $\Ju(\xv)=\mob^\prime(\phi)\hat\mu\grad\phi$ in $\vu{I}_\mathrm{bulk}$ scales as $\ell^{-3}$. Now, since $\Ju$ is a vector, isotropy imposes that it has zero average; we can then decompose $\vu{I}_\mathrm{bulk}$ into $(L/\ell)^d$ uncorrelated domains of volume $\ell^d$ where the integrand is of order $\ell^{-3}$. This yields a contribution to $\vu{I}$ that scales as:
\begin{equation}
|\vu{I}_\mathrm{bulk}|\lesssim L^{-d}\left(\frac{L}{\ell}\right)^{d/2}\ell^{d-3}\sim L^{-d/2}\ell^{d/2-3}.\label{eq:I_Ju_bulk}
\end{equation}

Turning now to the interface contribution in \eqref{eq:I_Ju2}, it is pedagogically useful to treat first the case where the system is not bicontinuous at late times, so that the minority phase comprises $N\sim (L/\ell)^d$ well separated droplets. (We do {\em not} assume these to be spherical.) Up to a dimensionless geometric factor set by their phase volume (itself set by $\bar\phi/\phi_B$), the separation and mean radius of the droplets both scale as $\ell(t)$. The interfacial term is then additive over droplets:
\begin{align}
\vu{I}_\mathrm{interfaces} & =  \frac{1}{L^d}\sum_{n=1}^N\vu{I}_n,\label{eq:I_Ju_drops_decomposition}
\end{align}
where $\vu{I}_n$ denotes the integral of $\Ju$ over the interfacial region of the $n$th droplet.

Next, we estimate the scaling of $\vu{I}_n$ for a given droplet (assumed far from the system boundary). We choose curvilinear coordinates comprising one coordinate $u_\perp$ perpendicular to the interface ({\em i.e.}, along $\vu{n}_\perp=\frac{\grad\phi}{|{\grad\phi}|}$), with $u_\perp = 0$ at the interface, and $d-1$ coordinates $\vb{s}$ tangential to the interface. Noting that the flux is everywhere perpendicular to the interface ($\vu{J}= \hat{J}_\perp \vu{n}_\perp$), we find:
\begin{align}
\vu{I}_n &= \int\dd[d-1]{\bf s}\,\vu{n}_\perp({\bf s})\int\dd{u_\perp}\mathcal{J}(u_\perp,\vb{s}) \hat{J}_\perp(u_\perp,{\bf s}).
\end{align}
Here the Jacobian of the transformation between Euclidean and $(u_\perp,{\bf s})$ coordinates is denoted by $\mathcal{J}(u_\perp,{\bf s})$. The integral over $u_\perp$ runs over a region of thickness given by the interfacial width $\sim\xi$. Furthermore, we note that the chemical potential is proportional to the local curvature to leading order, $\hat{\mu}\sim \div\vu{n}_\perp + O(\ell^{-2})$. Using a sharp interface limit, the integral reads:
\begin{equation}
\vu{I}_n = \int\dd[d-1]{\bf s}\,\vu{n}_\perp({\bf s}) \div\vu{n}_\perp({\bf s}) \mathcal{J}(u_\perp=0,{\bf s}) \hat{I}_\perp + O(\xi\ell^{d-3}), \label{eq:I_Ju_drops}
\end{equation}
where the integral along $u_\perp$ is denoted $\hat{I}_\perp\sim \xi$. At first sight, using that $ \div\vu{n}_\perp \sim \ell^{-1}$, this leading-order term scales as $\sim\xi\ell^{d-2}$. However,  to this order in $\ell$, $\hat{I}_\perp$ does not depend on the tangential coordinates $\vb{s}$, so that what remains in \eqref{eq:I_Ju_drops} is an integral of the mean curvature vector $(\div\vu{n}_\perp)\vu{n}_\perp$ over a smooth closed surface. This vanishes identically \cite{Blackmore1985}. Hence, the leading non-vanishing contribution scales as $\vu{I}_n \sim \xi\ell^{d-3}$. This object is a vector, whose direction is set by the asphericity and orientation of the $n$th droplet, and is therefore a random variable of mean zero in an isotropic system.

Now, $\vu{I}_\mathrm{interfaces}$ is a sum of these random vectors, which must be uncorrelated at scales beyond (some fixed multiple of) $\ell$. This gives the scaling
\begin{align}
    |\vu{I}_\mathrm{interfaces}| \lesssim L^{-d}\left(\frac{\ell}{L}\right)^{d/2}\xi\ell^{d-3}\sim L^{-d/2}\ell^{d/2-3},\label{eq:I_Ju_drops_scaling}
\end{align}
which is of the same order in $\ell$ and $L$ as the bulk contribution in \eqref{eq:I_Ju_bulk}.

In the estimate above, we have neglected the droplets that intersect the boundary of the system. For those droplets, the integral in \eqref{eq:I_Ju_drops} does not involve a closed surface, and hence need not vanish, implying $|\vu{I}_n|\sim \xi\ell^{d-2}$. At the boundaries of the system, the average of $\vu{I}_n$ at scales larger than $\ell$ will, by symmetry, be a vector of constant magnitude pointing along the boundary normal $\vu{n}$. This vector integrates to zero over the entire surface. Hence, only fluctuations about this average contribute to $\vu{I}$; these fluctuations are random zero-mean variables of magnitude $\sim\xi\ell^{d-2}$, correlated patchwise along the boundary. Analogously to the terms discussed in Sec.~\ref{sec:det-stress}, this results in a contribution to $\vu{I}$ of order $O(L^{-(d+1)/2}\ell^{(d-3)/2}\xi)$, which gives the same scaling seen there, and is harmless.

The argument above generalises rather directly to the case of bicontinuous domains rather than aspherical droplets. Any closed droplets can be treated as before, and where an extended interface cuts the system boundary this can also be closed off smoothly so that the integral over the now-closed surface, $\oint(\div\vu{n}_\perp)\vu{n}_\perp$ vanishes as before. This means that the leading-order term in $\vu{I}_\mathrm{interfaces}$ vanishes just as in \eqref{eq:I_Ju_drops}, leaving an integrand of order $O(\ell^{-2})$. The integrand in $\vu{I}_\mathrm{interfaces}$ is then a zero-mean random vector, correlated over patches of area $\ell^{d-1}$; since the total interface area in the system is of order $O(L^d/\ell)$, the total number of patches scales like $\sim(L/\ell)^d$. Each patch contributes a term that scales as $\sim\xi\ell^{d-1}\ell^{-2}$. This leads to the same scaling as in \eqref{eq:I_Ju_drops_scaling}. Finally, the contribution from the closures we have added at the boundary needs to be subtracted; analogously to the contribution found before for droplets that cross the boundary, this term scales as $O(L^{-(d+1)/2}\ell^{(d-3)/2}\xi)$.

We now analyse the implications of the above-found scalings for the hyperuniformity exponent $\qexp$. Integrating $\partial_t\rcom = \vb{I}$ gives a contribution $|\rcom| \lesssim \sqrt{S_0}L^{1-d/2}+L^{-d/2}\ell^{d/2-3+1/\nu}$, where the first term appears when initial conditions are not hyperuniform. Note that, in contrast to the constant mobility case, the part of $\rcom$ that scales as $L^{-d/2}$ is no longer conserved as time (or $\ell$) increases, which is a consequence of the appearance of a volume integral in $\vb{I}$.

Nonetheless, this scaling leads to $\partial_t S_2\lesssim L^d |\rcom|\cdot|\vb{I}| \lesssim \sqrt{S_0}L^{1}\ell^{d/2-3} + \ell^{d-6+1/\nu}$, which in turn yields $s_2 \lesssim S_0^{1/2}L^1\ell^{-(d/2+5)+1/\nu} + \ell^{2(1/\nu-4)}$. Once again using the causality length $L_\mathrm{eff}$ as an effective system size, this bound sharpens to:
\begin{equation}
s_2 \lesssim S_0^{1/2}\ell^{-(d/2+5)+(1+\gamma)/\nu} + \ell^{2(1/\nu-4)}.
\end{equation}
For $\nu=1/3$ and $\gamma=1/2$, the first term, coming from non-hyperuniform initial conditions, decays in time as $\lesssim S_0^{1/2}\ell^{-(d+1)/2}$ for any dimension (and more strongly if one argues $\gamma < 1/2$). The second term, on the other hand, decays as $\ell^{-2}$. We conclude that even with $\phi$-dependent mobility the hyperuniformity exponent reads $\varsigma=4$. However, this outcome is no longer directly attributable to $\rcom$-conservation, which is why a significantly more involved argument was required to produce a suitable bound on $s_2$.

\begin{figure}
    \centering
    \includegraphics[width=\textwidth]{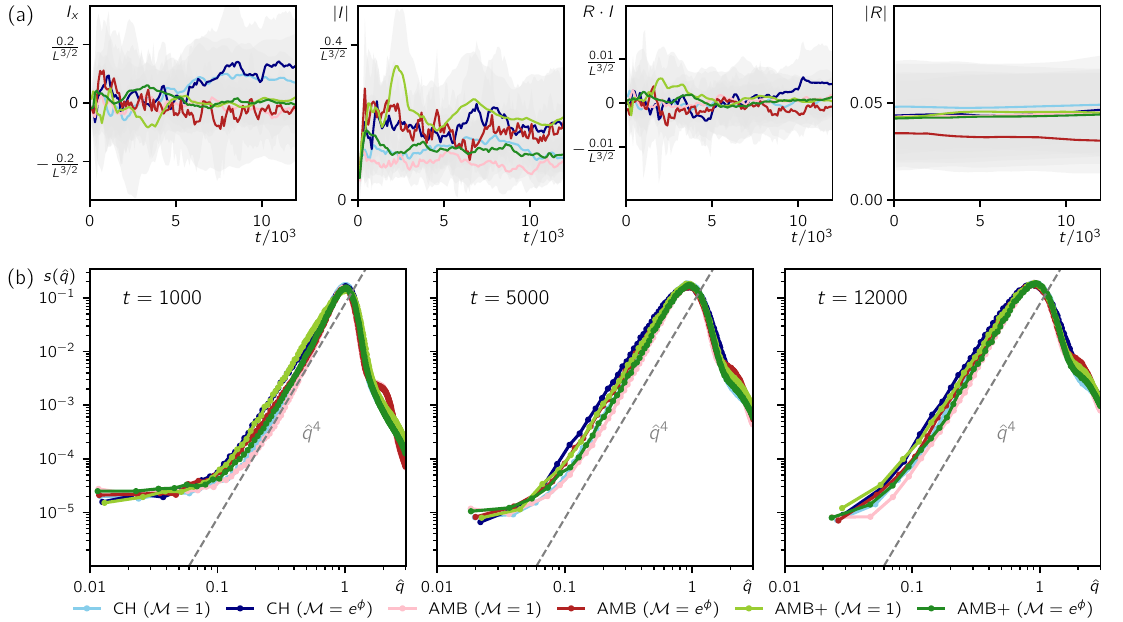}
    \caption{Hyperuniformity in deterministic models of coarsening. \textbf{(a)} First two panels: Time evolution of the $x$-component and the norm of  $\vb{I}$ defined in \eqref{eq:I}, confirming the scaling $\abs{I}\sim L^{-(d+1)/2}$ predicted in Sec.~\ref{sec:det-stress}. Last two panels: Time evolution of $\rcom\vdot\vb{I}=\partial_t\rcom^2/2$ and the norm of the centre of mass vector $\rcom$, showing that $|\rcom|$ conserves its initial $O(S_0^{1/2}L^{1-d/2})$ value in all models. Solid lines represent means over 10 runs, while standard deviations are shown in grey. \textbf{(b)} Rescaled structure factor $\hat{s}(\qh)=S(\qh \ell)/\ell^2$ at three different times. On the left of the peak, the constant and quadratic parts from the initial conditions disappear over time, leaving the predicted $\qh^4$ scaling for all models. Each dataset is averaged over 10 runs.} \label{fig:det}
\end{figure}

\subsection{Numerical simulations}\label{sec:2-num}
To test our theoretical results, we perform finite-difference numerical simulations of the two-dimensional Cahn-Hilliard equation, and noiseless Active Models B and B+, in a $2048\cross 2048$ geometry, with unit lattice spacing $\Delta x = \Delta y = 1$ and timestep $\Delta t = 0.01$, using the algorithm presented in \cite{Tjhung2018}. We modified the algorithm to impose no-flux boundary conditions, $\Jv\vdot\vu{n} = 0$, implementing Neumann boundary conditions for the field, $\grad\phi\vdot\vu{n} = 0$, see \ref{app:Numerics} for details. We prepare the system in the homogeneous state $\phi = 0$ and add a perturbation to each lattice point which is drawn from a uniform distribution over $[-0.2,0.2]$; we then re-scale the density globally so that $\bar\phi = 0$ holds exactly. We fix the parameters $K = 1$, $a = b = 0.25$, and choose $\lambda = 1.0$ and $\zeta = 0$ for Active Model B and $\lambda = 0$ and $\zeta = 2.0$ for Active Model B+. For the constant mobility case, we keep $\mob = 1$, while we choose $\mob = e^\phi$ to exemplify the non-constant case. Note that our initial conditions imply $S_0\sim O(1)$ and $S_{n>0} = 0$.

In Fig.~\ref{fig:det}a, we show
the $x$-component of $\vb{I}$ and its norm for the various models, as well as $\rcom\vdot\vb{I}$ and the norm of $\rcom$. The curves are obtained after averaging over 10 simulation runs that differ in the noise in the initial conditions. As shown, these quantities are comparable with $\vb{I}\sim O(L^{-3/2})$, which is the predicted scaling for $d=2$ for the case where $\Jv=-\div\Sv$ discussed in Section \ref{sec:det-stress}. This fully confirms our prediction that $\partial_t \rcom$ is negligible: indeed, $|\rcom|$ stays close to its initial value (set by the noise in the initial conditions), which for $d=2$ is of order $O(S_0^{1/2}L^0)$. Moreover, for non-constant mobility (where $\Ju\neq0$) the observed scaling means that $|\vb{I}|$ goes to zero at large $L$ much faster than the bound we derived in Sec.~\ref{sec:deterministic-mob}, suggesting that the bound in question might be tightened further by applying more mechanistic insight beyond that currently encoded in our generalised Tomita-like arguments.

This confirms empirically, but in full accord with our mechanistic discussion, that quadratic scaling in the structure factor should be suppressed in all models considered here. To rescale the structure factor extracted from simulations, we define the characteristic length scale $\ell$ in the following way:
\begin{equation}
\label{eq:qell}
	\frac{2\pi}{\ell} = q_\ell = \frac{\int_0^\infty\dd{q}qS(q)}{\int_0^\infty\dd{q}S(q)}.
\end{equation}
Using \eqref{eq:qell}, in Fig.~\ref{fig:det}b we plot the rescaled structure factor $s(\qh) = S(q)/\ell^2$ against the scaling variable $\qh = q/q_\ell$. For a large range of wave vectors left of the peak, all the models considered here exhibit the expected long-wavelength $s(\qh)\sim\qh^4$ scaling. These results fully confirm our theoretical predictions that (a) for constant mobility the hyperuniformity exponent $\qexp = 4$ is valid for deterministic theories of phase ordering regardless of activity, as observed in earlier numerics~\cite{Ma2017,Zheng2023}; and (b) that this result extends to non-constant mobility as well both in Cahn-Hilliard and in (noiseless) AMB+. 

\subsection{Comments on the physical origin of centre-of-mass conservation}\label{sec:CoM}
Here, we briefly comment on the physical conditions under which one should expect the presence of a conservation law for the centre of mass of the phase separating species. For the case where $\phi$ represents the physical density of a single species ({\em e.g.}, a set of passive Brownian particles with attractive interactions), then the physical centre of mass $\rcom_\mathrm{phys}$ of the particle density $\rho$ differs from that of the order parameter only through the constant shift $\varphi = \rho-\bar\rho$, so that $\partial_t\rcom_\mathrm{phys}=\partial_t\rcom$. Therefore we shall drop the distinction between $\rcom_\mathrm{phys}$ and $\rcom$.
Now, if the deterministic dynamics conserve the centre of mass $\rcom$ exactly, this requires $\vb{I} = {\bf 0}$ in \eqref{eq:I}. In particular, centre of mass conservation implies $\Ju = {\bf 0}$, so that the mass flux can be written as $\Jv = -\div\Sv$. This applies generally, including to theories that couple the scalar field to additional order parameters.

For passive Brownian particles, Newton's third law is encoded in the free energy, whose existence implies $\delta\mu({\bf r})/\delta\phi({\bf r'})=\delta\mu({\bf r'})/\delta\phi({\bf r})$. But since in Model B and its relatives momentum is not conserved, for this dynamics to conserve $\rcom$ at deterministic level a constant $\mob$ is also required, so that the {\em velocities} induced by interparticle forces between any pair of particles are equal and opposite.
(This differs from the case of undamped Newtonian dynamics where the third law alone is enough to conserve global momentum, and thus conserve $\rcom$ in the zero-momentum frame.) This observation is consistent with the finding in Section \ref{sec:deterministic-mob} above that $\Ju = {\bf 0}$ requires constant mobility.  It is remarkable however that the minimal terms used to break detailed balance in Active Model B+ do not themselves create nonzero $\Ju$, even though there is no notion of the third law for these terms (see Section \ref{sec:det-stress} above). Therefore it seems likely that introducing active terms beyond the minimal ones chosen in AMB+ would cause $\rcom$-conservation to be lost, and {\em possibly} convert $\qexp = 4$ to $\qexp = 2$. But, as explored in Section \ref{sec:noise} below, noise in Model B or AMB+, which is an inherent part of these stochastic field theories, already breaks centre-of-mass conservation -- the microscopic interpretation being that it assigns random additional velocities to each particle which do not sum zero. 

Equation \eqref{eq:A_dt} shows that in systems with centre-of-mass conservation $\partial_t\A = 0$ exactly, implying that $\A$ is conserved. Hence, if the initial conditions are isotropic, and the structure factor grows in time overall, centre-of-mass conservation directly explains why the quadratic scaling of the structure factor is suppressed, giving $\qexp=4$. 
(This result is still valid if centre of mass conservation holds in the bulk but is broken by boundary interactions, since we have shown that a surface integral in $\vb{I}$ leads to a subdominant contribution to $\partial_t\A$.) Notably, systems that conserve the centre of mass microscopically must still do so after coarse-graining so long as both the deterministic and noise terms are constructed in a consistent way. Below in Sections \ref{sec:stoch} and \ref{sec:R-D} we consider models where the noise is indeed structured so as to preserve the conservation of $\rcom$.

Whether introduced deliberately or by default, the existence of the $\rcom$-conservation law offers a simple and precise explanation of the result $\qexp=4$ in any model for which  $\Jv = -\div\Sv$ for some local $\Sv$. However, the converse does not hold: $\qexp=4$ does not imply this conservation law. Indeed we have shown in Section \ref{sec:deterministic-mob} above that non-constant mobility breaks the conservation law, but does not restore $\qexp=2$. We shall next show that the same is true for noise in $d\ge2$ dimensions.

\section{The role of noise}\label{sec:noise}
In this Section, we describe the effect of dynamical noise on the hyperuniformity of phase-ordering systems. By extending the arguments given above for the deterministic case, we show that although the structure factor is still proportional to $q^4$ in the scaling regime, noise induces a transient in which $s(\qh)\sim \qh^2$ when $\qh\ll \qhcross\sim t^{[1-\nu(d+2)]/2}$ and $d\geq 2$. This is discussed in Sec.~\ref{sec:stoch}, where we also assess the expected crossover between quadratic and quartic scaling numerically in the case of Model B with constant $\mob$. 
In Sec.~\ref{sec:1-d}, we consider the one-dimensional case where the crossover is constant in time, so that $\qhcross$ vanishes only for $D\to 0$, not $t\to\infty$. In Sec.~\ref{sec:R-D} we consider a reaction-diffusion model that is exactly solvable, showing that the arguments developed in this paper remain valid beyond the scope of coarsening models. 

\subsection{Hyperuniformity in stochastic coarsening models}\label{sec:stoch}
\begin{figure}
    \centering
    \includegraphics[width=0.75\textwidth]{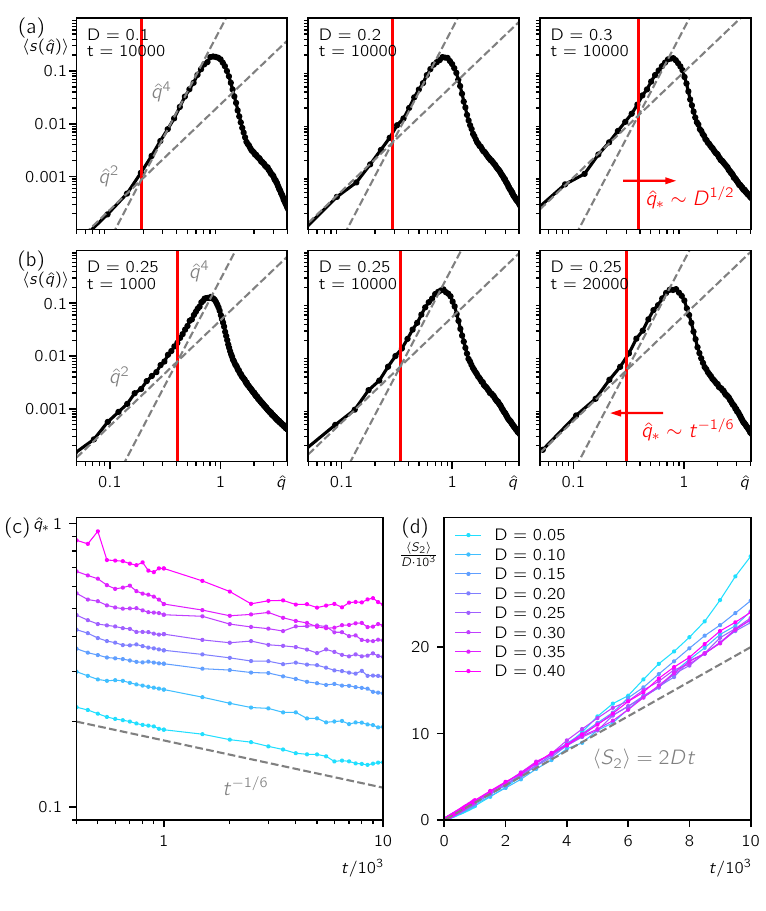}
    \caption{Hyperuniformity in Model B in $d=2$ with dynamical noise controlled by $D$. \textbf{(a)} Rescaled structure factor $\langle \hat{s}(q)\rangle=\langle S(q)\rangle /\ell^2$ averaged over 20 runs in a system with periodic boundaries of size $1024\cross 1024$. For finite noise strengths $D$, in addition to the $\qh^4$ scaling seen in Fig.~\ref{fig:det}, a $\qh^2$ scaling region appears for small wave vectors. The crossover wave vector $\qhcross$ moves to the right as $D$ is increased for fixed run-times $t$. \textbf{(b)} For fixed $D$, the crossover wavelength moves to the left as time progresses.
    \textbf{(c)} Measurements of $\qhcross$ averaged over 20 runs following the procedure detailed in the main text reveal a slow power-law decay which is  compatible with the theoretically predicted $-1/6$ exponent (colours as in panel d). \textbf{(d)} Measurements of $\A$ as detailed in the main text, averaged over 20 runs, showing the noise-dependent growth of the quadratic coefficient.}
    \label{fig:stoch}
\end{figure}

We now study the effect of setting $D\neq0$ in \eqref{eq:continuity}. Broadly speaking, the noise $\Lv$ then results in a volume integral in \eqref{eq:I}, which is no longer localised to the interface between phases, thus breaking the arguments given in Sec.~\ref{sec:det-stress} above. For simplicity, as there, we restrict ourselves to the case of constant mobility $\mob$, and consider hyperuniform initial conditions, so that $S_0 = 0$. (This includes the simplest case of zero initial noise, which will now phase separate since $D\neq 0$.) With these restrictions we can rigorously show how dynamical noise affects the structure factor. 

We start by generalising \eqref{eq:A_dt} to the noisy case. The vector $\rcom$ is a linear function of the stochastic field $\df(\xv)$, so that its time evolution follows directly from \eqref{eq:continuity}. Integrating by parts as in \eqref{eq:I}, we obtain:
\begin{equation}
    \partial_t\rcom = \frac{1}{L^d}\int\dd[d]{\xv}\Jv(\xv) + \frac{1}{L^d}\int\dd[d]{\xv}\Lv(\xv) := \vb{I}_\mathrm{det} + \boldsymbol{\eta},\label{eq:dt_R_stoch}
\end{equation}
where $\vb{I}_\mathrm{det}$ is the deterministic contribution given in \eqref{eq:I}, while $\boldsymbol{\eta}(t)$ is a Gaussian random vector with zero average and variance:
\begin{equation}
\expval{\eta_i(t)\eta_j(t^\prime)} = G_{ij}\delta(t-t^\prime);\qquad G_{ij} = \frac{2D}{L^d}\delta_{ij}.
\end{equation}

Next we consider the noise-averaged structure factor, $\expval{S(q)}$. When $S_0$ can be neglected, the quadratic coefficient $S_2$ is a function of $\rcom$. Its time evolution follows from \eqref{eq:dt_R_stoch} by applying It\^o's lemma~\cite{Gardiner}:
\begin{equation}
    \expval{\partial_tS_2} = \expval{\pdv{S_2}{R_i} I^i_\mathrm{det}} + \expval{\pdv{S_2}{R_i}\eta_i} + \frac{1}{2}\pdv{S_2}{R_i}{R_j}G_{ij},\label{eq:Ito}
\end{equation}
where summation over repeated indices is implicit. Using the relation \eqref{eq:A_R} and $\delta_{ij}\delta_{ij}=d$, we obtain:
\begin{equation}
    \expval{\partial_tS_2} = \frac{2L^d}{d}\expval{\rcom\vdot\vb{I}_\mathrm{det}} + \frac{2L^d}{d}\expval{\rcom\vdot\boldsymbol{\eta}} + 2D.\label{eq:A_dt_noise}
\end{equation}
Here, the first term is the average deterministic time evolution given in \eqref{eq:A_dt}. The second term vanishes due to the non-anticipating property of the It\^o integral and from $\expval{\boldsymbol{\eta}(t)}={\bf 0}$. The last term in \eqref{eq:A_dt_noise} is a constant noise contribution. Being $O(L^0)$, for $D\neq 0$ this dominates the first term, which as discussed in Sec.~\ref{sec:det-stress} scales like $\sim L^{-1/2}$. We will neglect the deterministic contribution in the following.

Then, integrating \eqref{eq:A_dt_noise} yields:\begin{equation}
\label{eq:A_stoch}
    \expval{\A} \approx 2Dt.
\end{equation}
Thus, the volume term in \eqref{eq:dt_R_stoch} leads to a linear growth of the quadratic coefficient of the (noise-averaged) structure factor, in contrast to conservation of $S_2$ found for constant mobility without noise. Recalling \eqref{eq:rescaled_S_coeff}, this gives for the rescaled structure factor:
\begin{equation}
\label{eq:s2_t}
    \langle s_2\rangle \sim Dt\cdot\ell^{-(d+2)} \sim Dt^{1-\nu(d+2)},
\end{equation}
where we have neglected the dependence of $\ell$ on $D$, which is justified for small $D$ and $d\geq 2$ (see below for the $d=1$ case).

This result implies a crossover from a quadratic to a quartic regime in $\langle s(\qh)\rangle$ taking place at a rescaled wave number $\qhcross$. Assuming $\langle s_4\rangle$ constant in time and independent of $D$, we find:
\begin{equation}\label{eq:qstar-C}
    \qhcross \sim \sqrt{D} t^{[1-(d+2)\nu]/2}\,.
\end{equation}
As long as $\nu (d+2)>1$, while the noise leads to a growing quadratic coefficient $\A$, the resulting contribution to the rescaled structure factor $\langle s(\qh)\rangle$ does not survive in the self-similar scaling regime, as \eqref{eq:s2_t} predicts $\expval{s_2}$ to decay in time. In passive fluids, it is well known that $\nu=1/3$ and thus, for dimension $d\geq 2$, $\qhcross \sim \sqrt{D}t^{-(d-1)/6}$. Therefore, the rescaled structure factor becomes universal (independent of $D$) at late times where $\qhcross\to 0$ and pure 
$s(\qh)\sim\qh^4$ scaling emerges. Whereas for active field theories the precise $\nu$ exponent remains debated~\cite{Wittkowski2014,pattanayak2021ordering}, we have not seen clear evidence of it being sufficiently different from $\nu = 1/3$ to change this outcome. 

We now test our predictions concerning the crossover by performing finite-difference numerical simulations of passive Model B in a $1024\cross 1024$ geometry (with periodic boundary conditions for simplicity), starting from a homogeneous initial state with $\phi = 0$. We use the same parameters as for the deterministic simulations presented earlier, and vary the noise strength $D$.
Figure~\ref{fig:stoch}a shows the rescaled structure factor curves at equal times for different values of $D$, averaged over 20 runs. In contrast to the deterministic case (see Fig.~\ref{fig:det}b), a $\qh^2$ scaling regime is now present at small $\qh$. As expected, the crossover between this regime and the quartic scaling shifts to the right with increasing noise strength $D$. Likewise, Fig.~\ref{fig:stoch}b shows $\langle s(\qh)\rangle$ at fixed $D$ for different times, which confirms the shift of the crossover wave vector $\qhcross$ to the left when time increases, as predicted by \eqref{eq:qstar-C}.

To extract the crossover quantitatively, we rewrite the structure factor as
\begin{equation}
\label{eq:S_fit}
    S(q) = \A q^2\left[1+\left(\frac{q}{\qcross}\right)^2\right] + O(q^6).
\end{equation}
With $y = S(q)$ and $x = \log q$, this becomes:
\begin{equation}
\label{eq:fit}
    y = \alpha + 2x + \log(1+\beta e^{2x}),
\end{equation}
where $\A = e^\alpha$ and $\qcross = \beta^{-1/2}$. We then fit the parameters $\alpha$ and $\beta$ in \eqref{eq:fit} to the structure factor data for every time step, obtaining $\A(t)$ and $\qcross(t)$. We apply this procedure for each run and then average the result over $20$ runs. In Figure~\ref{fig:stoch}c, we show the resulting $\qhcross(t)$, exhibiting a clear power-law decay, consistent with our theoretical prediction $\qhcross\sim t^{-1/6}$ for $d=2$ and $\nu=1/3$. Furthermore, rescaling the $y$-axis by $\sqrt{D}$ results in curve collapse for small values of $D$ (not shown), as expected from \eqref{eq:qstar-C}.

Finally, in Fig.~\ref{fig:stoch}d we report the linear growth of $\A$ with time, confirming that \eqref{eq:A_stoch} holds. The deviations from the theoretical expectation seen at late times might be because the quadratic regime is suppressed, so that our fitting procedure becomes less accurate at late times, which is especially true at small noise strengths.

In conclusion, we have shown how Tomita's argument can be extended to models of phase separation including dynamical noise. The noise induces a new regime in the rescaled structure factor, which now goes to zero as $s(\qh)\sim\qh^2$ for $\qh\ll \qhcross$, with the crossover wavenumber $\qhcross$ converging to zero slowly with time when $d\geq2$. We have not addressed the case of activity, but since in AMB+ the noise breaks the $\rcom$ conservation law in a similar way, we expect a similar outcome. We also assumed constant mobility $\mob$; without this, the conservation law is already broken (albeit without changing the result $\qexp=4$) and a more complicated theoretical argument, extending that of Section \ref {sec:deterministic-mob} above to include noise, is needed. This lies beyond the scope of the current paper.

\subsection{The one-dimensional case}\label{sec:1-d}
\begin{figure}
    \centering
    \includegraphics[width=\textwidth]{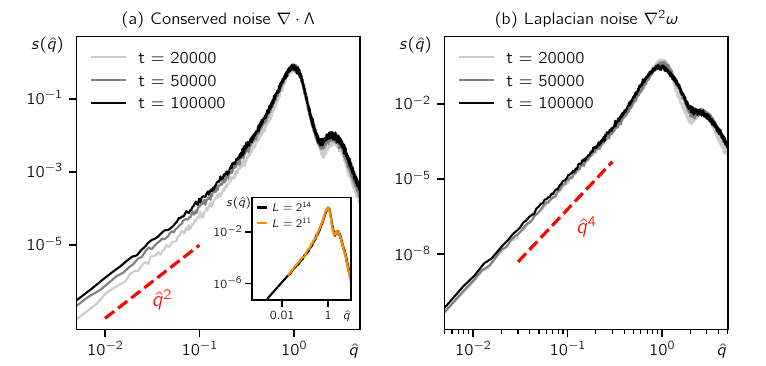}
    \caption{Rescaled structure factor $\langle s(\qh)\rangle$ for $d=1$ with standard conserved noise (Model B), and with Laplacian noise. \textbf{(a)} $\langle s(\qh)\rangle$ for Model B in a system of size $L=2^{14}$ with periodic boundary conditions, averaged over 100 runs. Noise strength $D=10^{-4}$. The crossover wave vector $\qhcross$ is now time-independent, so that for $D\neq 0$ the hyperuniformity exponent is $\qexp = 2$, not $\qexp=4$. Inset: $\langle s(\qh)\rangle$ at $t=10^5$ for different system sizes $L$. The crossover point $\qhcross$ does not change with $L$. \textbf{(b)} Using Laplacian noise instead of the standard Model B noise drastically alters the long-wavelength behaviour of the structure factor, resulting in the suppression of the quadratic regime and recovering $\varsigma = 4$ as holds in the case of deterministic dynamics ({\em i.e.}, at $D=0$). 
    }
    \label{fig:1d}
\end{figure}
The results of the previous Section are valid for $d\geq 2$. The case $d=1$ is special: here, interfaces are domain walls without curvature, so that domain growth is no longer driven by curvature-driven (Ostwald) dynamics. In consequence, without noise, coarsening is drastically slowed down, with $\ell$ showing sub-power-law behaviour, instead growing as $\ell \sim \log t$. However, when noise is introduced, a new mechanism of diffusive coalescence drives coarsening and restores power-law growth, which for Model B has the usual $\nu = 1/3$ exponent, with $\ell \sim (Dt)^{1/3}$ \cite{Kawakatsu1985, Bray1994,Majumdar1995,Cates2018}.

A surprise however arises in the small wave number behaviour of the structure factor: Inspecting \eqref{eq:s2_t} for $d=1$ and using $\nu = 1/3$, we see that the exponent on the right hand side vanishes. Hence, for the noisy $d=1$ case, $\langle s_2\rangle $ is time-independent, resulting in a hyperuniformity exponent $\qexp=2$ in the large $t$ scaling limit, in contrast to $\qexp = 4$ for the deterministic case. The resulting $D$-dependence of $\qh$ is also interesting. With $\ell\sim (Dt)^{1/3}$ one has $s_2 \sim D^0$; on the other hand, since $S_4\neq 0$ for the deterministic case, we can assume $S_4\sim D^0$ at least in the low-noise regime, so that \eqref{eq:rescaled_S_coeff} gives $s_4 \sim D^{-5/3}$ and we obtain $\qhcross\sim D^{5/6}t^0$. This is therefore not merely a special case of \eqref{eq:qstar-C}, according to which
$\qhcross\sim D^{1/2}t^{[1-(d+2)\nu]/2}$ for $d\ge 2$. 

The noise-induced change in hyperuniformity exponent from $\qexp = 4$ to $\qexp=2$ is crucially dependent on the form of the noise used. Although the usual Model B noise
\eqref{eq:Lambda_var} is the physically relevant one in most situations, one might instead  choose a noise of Laplacian form, so that the one-dimensional equations read:
\begin{equation}
\label{eq:1d_lap}
    \partial_t\phi = -\partial_xJ_x + \partial^2_x\omega,
\end{equation}
where $\omega$ is a Gaussian random variable with variance: \label{eq:omeganoise}
\begin{equation}
    \expval{\omega(x,t)\omega(y,t^\prime)} = 2D\delta(x-y)\delta(t-t^\prime).
\end{equation}
Importantly, a noise of this form results in a stochastic contribution to the second-rank tensor $\Sv$ in \eqref{eq:J_decomposition} rather than to $\Ju$. Therefore, unlike standard Model B noise, Laplacian noise {\em maintains the conservation law for the centre-of-mass vector}, $\rcom$ \cite{Hexner2017}, in which case the $\qexp = 4$ exponent ought to be unaffected. More precisely,
reproducing the steps that led previously to \eqref{eq:A_dt_noise}, we find that Laplacian noise adds an additional gradient to the integral over noise in \eqref{eq:dt_R_stoch}, which can then be written as a surface integral. As a consequence of this, the term proportional to $D$ in \eqref{eq:A_dt_noise} no longer scales $O(L^0)$, but instead becomes  $O(L^{-1})$~\footnote{The coefficient of the $O(L^{-1})$ part diverges for strictly $\delta$-correlated noise as in \eqref{eq:omeganoise} but a finite noise correlation length can instead be assumed for the purposes of this and similar arguments; see {\em e.g.} \cite{cates2022entropy}.}. Therefore, Laplacian noise effectively conserves $\langle \A\rangle$, from which we conclude that $\langle s(\qh)\rangle\sim \qh^4$, as in the deterministic $d=1$ case.

To verify the above predictions, we perform numerical simulations of Model B in $d=1$ with standard noise as well as Laplacian noise, using the same parameters as for the $d=2$  simulations presented above, and fixing the  noise strength to $D=10^{-4}$. The resulting rescaled structure factors $\langle s(\qh)\rangle $ are shown in Fig.~\ref{fig:1d}. Here, instead of the definition \eqref{eq:qell}, we have averaged $S(q)$ over batches of 10 runs and then defined $q_\ell$ as the position of the maximum. Rescaling the structure factor with this $q_\ell$, we then averaged it over 10 batches, {\em i.e.}, 100 runs in total. In agreement with our theory, the rescaled structure factor shows an extensive region of quadratic scaling at small wave vectors for standard Model B noise which does not vanish with time, in contrast to the two-dimensional case ({\em cf.} Fig.~\ref{fig:stoch}b and c). We conclude that the hyperuniformity exponent remains $\qexp=2$ indefinitely for one-dimensional coarsening systems with Model B-type noise~\footnote{Technically we have shown this only for the passive, constant mobility case -- but generalising to include activity and non-constant $\mob$ seems very unlikely to {\em remove} the $s_2$ term. Indeed in higher dimensions, where $s_2$ is absent in the same special case, the challenge addressed in Section \ref{sec:det} was only to show whether the additional terms could {\em reinstate} it.}.
Conversely, using Laplacian noise as in \eqref{eq:1d_lap} results in the suppression of quadratic scaling, leaving quartic scaling as the lowest non-vanishing order in the structure factor, so that the hyperuniformity exponent is $\qexp=4$.

\subsection{Comparator study: hyperuniformity in reaction-diffusion models}\label{sec:R-D}
We now discuss how hyperuniformity  arises in a different system, namely in a reaction-diffusion model. Our generalisation of Tomita's arguments also apply to this case, but in addition here the structure factor can be computed in closed form, allowing us to check the validity of such arguments directly. Building on \cite{Hexner2017,Ma2023}, this allows us to elucidate the role played by $\rcom$ conservation, and to analytically show the effects of the type of noise on the hyperuniformity exponent $\qexp$. 

The model consists of two species, active and passive, that live on a lattice \cite{Ma2023}. The active particles $A$ diffuse isotropically with diffusivity $\alpha$, whereas the passive particles $P$ are static. In addition to the diffusive dynamics, two reactions take place: an on-site activation reaction $A + P\to 2A$ with rate $\kappa$, whereby passive particles can convert into active ones in the presence of active particles, and a deactivation reaction $A \to P$ with rate $\mu$, which is independent of the particle's surroundings. We denote by $\rho(\xv)$ and $\rho_A(\xv)$, respectively, the total density and the density of active particles. 
This model has an absorbing-state phase transition: when the average density $\rho_0<p_0=\mu/\kappa$, all particles are passive in the steady state. In contrast, when $\rho_0>p_0$,  there remains forever a finite fraction of active particles in the thermodynamic limit $L\to \infty$. Like other systems possessing an infinite number of distinct absorbing states, this transition is thought to belong to the conserved Directed Percolation (C-DP) universality class, whose properties have been studied at length in the literature~\cite{VanWijland2002,LeDoussal2015,Janssen2016,Ma2023}. 

Here however we are not concerned with the critical regime, and rather consider the model deep in the active phase, far away from the C-DP phase transition. Under this premise, we may expand the coarse-grained dynamics to linear level, to give:
\begin{subequations}
\label{eq:C-DP}
\begin{align}
    \partial_t\rho_A &= \alpha\laplacian\rho_A + r(\rho-\rho_A-p_0)+\sqrt{2T}\tilde\omega + \sqrt{2D}\div\vb{\tilde\Lambda},\\
    \partial_t\rho &= \alpha\laplacian\rho_A + \sqrt{2D}\div\vb{\tilde\Lambda},
\end{align}
\end{subequations}
where $\tilde\omega$ and $\vb{\tilde\Lambda}$ are unit Gaussian white noises arising from birth/death (`demographic noise') and from random currents (`diffusive noise') respectively. In \eqref{eq:C-DP}, we introduce notation $T = \mu a_0$, $D = \alpha a_0$ and $r = \kappa a_0$, with $a_0 = \rho_0-p_0$. 

Note that the total density $\rho$ is a conserved field with diffusive noise whereas $\rho_A$ is not conserved; the active density dynamics includes an additional non-conservative noise $\tilde\omega$ that arises due to the reactions. Moreover it is simple to see that (since we are deep in the active phase) the fluctuations in the {\em passive} density $\df_P = \rho-\rho_A-p_0$ are fast. Indeed, taking the difference of the two equations in \eqref{eq:C-DP} we obtain
\begin{align}
    \partial_t\df_P &= -r\df_P-\sqrt{2T}\tilde\omega\,.
\end{align}
Adiabatically eliminating $\df_P$ and inserting the result into \eqref{eq:C-DP} gives:
\begin{align}
\label{eq:C-DP_adiab}
    \partial_t\rho &= \alpha\laplacian\rho + \laplacian\omega + \div\Lv
\end{align}
where $\omega = \sqrt{2T}(\alpha/r)\tilde\omega$ and $\Lv = \sqrt{2D}\vb{\tilde\Lambda}$. This model thus provides an example for the emergence of Laplacian noise from the elimination of a fast variable with demographic noise. The $\omega$ term in \eqref{eq:C-DP_adiab} is precisely the Laplacian noise considered in \eqref{eq:1d_lap} above.

The relation derived in \eqref{eq:A_R} between the second derivative of the structure factor $\A$ and the centre of mass vector $\rcom$ is generic for any scalar field, and hence still applies to \eqref{eq:C-DP_adiab}. This means that for isotropic and hyperuniform initial conditions, with $\rcom^2$ negligible, the initial structure factor obeys $S(q,0) \sim q^4$. Furthermore, since \eqref{eq:C-DP_adiab} has the form of a continuity equation, all arguments presented for the coarsening case are still valid here. Thus we expect that the deterministic part and the Laplacian noise $\omega$ should together lead to quartic scaling of the structure factor, $\qexp=4$, whereas the Model-B type conserved noise $\Lv$ should convert this to quadratic scaling, $\qexp=2$ \cite{Hexner2017}.  

Importantly, these arguments apply only for `intermediate' times, which we define as those short compared to that required to reach a stationary state. (This is also true of the scaling regime in the phase ordering problem, and in both cases the time scale required for stationarity diverges with the system size, $L$.) Note that once stationarity is reached, quadratic scaling is recovered even for $\Lv={\bf 0}$, but for \eqref{eq:C-DP_adiab} this emerges in a singular fashion whereby $S_2(t<\infty)=0$ while $S_2(\infty)=T$
\cite{Hexner2017}. (The solution in \cite{Hexner2017} is effectively $S(q) \sim Tq^2(1-e^{-2\alpha q^2t})$, which illustrates this singular property.)

To confirm the above expectations concerning the effect of noise-type on the hyperuniformity exponent $\qexp$, we derive in \ref{app:C-DP} exact results for the full two-component linearised model \eqref{eq:C-DP}, without resorting to the adiabatic elimination of $\varphi_P$. The resulting time-dependent structure factor of the total density $\rho$ is:
\begin{align}
    S(q,t) &=
    \frac{\alpha q^2 T}{(r-\alpha q^2)^2}\left[(1-e^{-2\alpha q^2t})+\frac{\alpha q^2}{r}(1-e^{-2rt})-\frac{4\alpha q^2}{(r+\alpha q^2)}(1-e^{-(\alpha q^2+r)t})\right]\nonumber\\
    &\phantom{{}={}}+\frac{D}{\alpha}(1-e^{-2\alpha q^2t})+\left[c_1(q)e^{-\alpha q^2t}+c_2(q)\alpha q^2e^{-rt}\right]^2.\label{eq:C-DP_S}
\end{align}
At $t=0$, this reduces to:
\begin{align}
\label{eq:C-DP_S0}
    S(q,0) &= \left[c_1(q)+c_2(q)\alpha q^2\right]^2,
\end{align}
where $c_1(q), c_2(q)$ are determined from the initial conditions for $\rho_A$ and $\rho$.

We now expand \eqref{eq:C-DP_S} for intermediate times, $t\ll L^2/\alpha$, and assume that $r \sim \alpha L^{-2}$. (This assumption is partly relaxed in \ref{app:C-DP}.) The latter implies that the reactive and diffusive times at the system scale are comparable or, equivalently, that we are considering systems of size comparable to the correlation length $\sqrt{\alpha/r}$ \cite{Ma2023}. With these assumptions, we find for small $q$:
\begin{equation}
\label{eq:C-DP_S_smallt}
    S(q,t) = S(q,0) + 2Dq^2t - 2S(q,0)^{1/2}\,[c_1(q)+c_2(q)r]\alpha q^2t + O((\alpha q^2t)^2).
\end{equation}
For initial conditions with $S_0(0)=S_2(0) = 0$, one has $c_1(q)\sim q^2$ and $c_2(q)\sim q^0$ in \eqref{eq:C-DP_S0}. This, in turn, makes the third term in \eqref{eq:C-DP_S_smallt} scale like $q^4$ (because $S(q,0)^{1/2}\sim q^2$). 

The result \eqref{eq:C-DP_S_smallt} shows that, in the absence of standard Model B noise ($D = 0$), the quartic scaling of the low-$q$ structure factor is conserved by the dynamics at intermediate times, so that $\qexp=4$. However, the term in $S(q,t)$ proportional to $D$ scales as $q^2$, so that Model B noise disrupts the initial quartic hyperuniformity and leads to the emergence of $\qexp = 2$ scaling. In contrast, the Laplacian noise term $T$ does not enter \eqref{eq:C-DP_S_smallt} and hence does not influence the dynamics of the structure factor in the time window under consideration here.
This exact calculation confirms the found by the general arguments applied above to the adiabatically reduced model \eqref{eq:C-DP_adiab}.

As discussed elsewhere \cite{Hexner2017,Ma2023} the choice of noise in \eqref{eq:C-DP} is deeply related to the presence or absence of microscopic centre-of-mass conservation among the interacting active and passive particles. Specifically, the Laplacian noise respects this conservation law whereas the Model B noise does not \cite{Hexner2017}. At the microscopic level one can choose a variant of the standard dynamics in which active particles only move pairwise with equal and opposite displacements, enforcing the conservation law. In the stationary state (rather than for intermediate times as considered above) this gives hyperuniformity throughout the active phase, with $\qexp=2$, whereas without centre-of-mass conservation the active phase is not hyperuniform at all, except at the C-DP critical point \cite{Hexner2017}. It seems likely that the same distinction also gives two sub-classes with different critical hyperuniformity exponents at the C-DP transition, a scenario explored in \cite{Ma2023}.

\section{Discussion}
\label{sec:discussion}

In this work, we have investigated the emergence of hyperuniformity in diffusive phase ordering, for a range of scalar field theories describing both active and passive fluids without momentum conservation. We did so by re-working and extending an analytical approach, developed by Tomita for the Cahn-Hilliard equation with constant mobility~\cite{Tomita1991}, to incorporate the effect of activity, state-dependent mobility, and noise. Central to this approach is the idea that in non-critical, hyperuniform systems, the coefficient of the lowest order ($\qh^2$) term in the rescaled structure factor $s(\qh)$ can be suppressed to zero by appeal to the system's isotropy at large scales, such that the hyperuniformity exponent $\qexp$ becomes $4$ rather than $2$. In cases without state-dependent mobility or noise, but including activity, we established that $\qexp=4$ directly by exploiting an explicit conservation law on the centre of mass vector $\rcom$ such that the time dependence of this quantity can be written as an integral over boundary terms at the edges of the system, which can then be shown negligible. For cases with non-constant mobility and/or noise there is no such strict conservation law, but by carefully estimating corrections to this picture we showed the value $\qexp = 4$ to be robust, with the exception of noisy systems in one dimension where $\qexp = 2$ is restored instead.

Thus, in the deterministic case, we showed that the rescaled structure factor goes as $s(\qh)\sim \qh^4$ for small $\qh$ for both passive and active models regardless of whether a constant or field-dependent mobility is chosen, and we numerically confirmed our predictions in two spatial dimensions. Our results agree with earlier numerical findings covering the Cahn-Hilliard equations \cite{Ma2017,Wilken2023} and active phase-separating fluids~\cite{Zheng2023} for constant mobility, while providing a theoretical explanation for the wider ubiquity of quartic hyperuniform scaling, $\qexp = 4$, in deterministic scalar field theories undergoing phase ordering.

Taking into account dynamical noise has instead a more subtle effect, as was shown in~\cite{Furukawa1989} (using different techniques from ours) for the Cahn-Hilliard equation with constant mobility. Our analysis confirms that $s(\qh)$ now acquires a $\qh^2$ contribution for $\qh\ll \qhcross$, and crossovers to $\qh^4$ only at larger $\qh$ values. We have shown that this crossover takes place at $\qhcross \sim \sqrt{D} t^{[1-(d+2)\nu]/2}$, where $\nu$ is the coarsening exponent. Assuming $\nu=1/3$, we obtained that $\qhcross$ goes to zero at large times for $d\geq 2$, so that, even with noise, a universal $\qh^4$ law is recovered in the scaling limit. On the other hand for $d=1$, $\qh_*$ saturates to a finite value, so that $\varsigma = 4$ in the deterministic case and $\varsigma = 2$ in the stochastic case. This mirrors experimental results in one dimension, which found different scalings of the structure factor depending on the relevance of thermal fluctuations \cite{Nagaya2004,Basu2016}, as well as theoretical studies on the 1d Ising model with Kawasaki dynamics \cite{Majumdar1994}. We showed that this crossover crucially depends on the form of noise, recovering quartic scaling if ($\rcom$-conserving) Laplacian noise is used instead of standard Model B noise.

The slow decay in time of the crossover wave number $\qhcross$ for $d\geq 2$ is notable. The unavoidable presence of noise at continuum level might explain why a long-lived $\qh^2$ scaling was observed experimentally in some active systems~\cite{Wilken2023}, whose noise levels tend to be larger than for their passive counterparts. Note also that our prediction that $\qhcross\to0$ for large times requires $\nu >1/(d+2)$. If, as has been suggested, $\nu < 1/3$ for active phase separation \cite{pattanayak2021ordering}, then $\qh_*$ might not vanish in $d\ge 2$ with the surprising consequence that active phase separation is then not governed by a zero-temperature fixed point as it is in passive case~\cite{Bray1994}. However, to reach this conclusion would require stronger numerical evidence for $\nu\neq 1/3$ than currently available.

Our theory provides a framework to study hyperuniform scaling in generic systems beyond models of coarsening of a single scalar field. We demonstrated this by developing our arguments for a reaction-diffusion model \cite{Ma2023}, where quadratic scaling in $\qh$ emerges in the intermediate, pre-steady-state time regime when standard Model B noise is used, but does not emerge if only Laplacian noise is present. (We showed this to be true in all dimensions, not just $d=1$ as was found for phase separation.) Furthermore, while our treatment predicts the suppression of quadratic scaling for deterministic theories with a single scalar field, an interesting question for further research is how this scaling is affected when such a scalar is coupled to additional fields, such as polar or nematic order parameters.

{\em Acknowledgements:} We thank Ronojoy Adhikari, Balázs Németh, Sriram Ramaswamy and Johannes Pausch for useful discussions. FDL acknowledges the support of the University of Cambridge Harding Distinguished Postgraduate Scholars Programme. XM thanks
the Cambridge Commonwealth, European and International
Trust and China Scholarship Council for a joint
studentship. CN acknowledges the support of the Institut National de Physique (INP) through the International Research Program (IRP) ``IFAM''. Work funded in part by the ANR grant ``PSAM''.
\\

\pagebreak
\appendix
\section{Numerical details on no-flux and Neumann boundary conditions}\label{app:Numerics}
In this Appendix, we present the implementation of no-flux conditions in the numerical simulations of the deterministic models of Sec.~\ref{sec:det}. We use a forward Euler method with a central finite-difference scheme for spatial derivatives. The first derivative of $\phi$ is discretised as:
\begin{equation}
\label{eq:firstdev}
\partial_x\phi_{i,j} = \frac{\phi_{i+1,j} - \phi_{i-1,j}}{2\Delta},
\end{equation}
and analogously for $\partial_y\phi$. At the boundary, we impose the Neumann condition $\grad\phi\vdot\vu{n} = 0$, which translates into $\partial_x\phi_{N,j} = 0$ and $\partial_x\phi_{0,j} = 0$. Hence, we calculate \eqref{eq:firstdev} using the two ghost points:
\begin{equation}
    \phi_{N,j} = \phi_{N-2,j}, \qquad\phi_{-1,j} = \phi_{1,j}.
\end{equation}
This derivative is used both for $\phi$ in the active terms \eqref{eq:Jact} and for $\mu$ in \eqref{eq:Jpass}, which implements the no-flux condition $\Jv\vdot\vu{n} = 0$.
The Laplacian is given by:
\begin{equation}
\Delta x\Delta y\laplacian\phi_{i,j} = \phi_{i+1,j} + \phi_{i-1,j} + \phi_{i,j+1} + \phi_{i,j-1} - 4\phi_{i,j},
\end{equation}
where the Neumann condition $\grad\phi\vdot\vu{n} = 0$ imposes:
\begin{equation}
    \phi_{N,j} = \phi_{N-1,j},\qquad \phi_{-1,j} = \phi_{0,j},
\end{equation}
and analogously for the $y$ coordinate. Finally, using \eqref{eq:continuity} and the derivative \eqref{eq:firstdev}, centre of mass conservation requires:
\begin{align}
0=\partial_t\sum_{i,j}\phi_{i,j} &= -\sum_{i,j}(\partial_xJ^x_{i,j}+\partial_yJ^y_{i,j}),\nonumber\\
 &= -\frac{J^x_{N,j}+J^x_{N-1,j}-(J^x_{0,j}+J^x_{-1,j})}{2\Delta x}\nonumber\\
 &\phantom{{}={}}-\frac{J^y_{i,N}+J^y_{i,N-1}-(J^y_{i,0}+J^y_{i,-1})}{2\Delta y},
\end{align}
leading to the condition on the ghost points:
\begin{equation}
    J^x_{N,j} = -J^x_{N-1,j} = 0,\qquad J^x_{-1,j} = -J^x_{0,j} = 0,
\end{equation}
and analogously for the $y$ coordinate.

\section{Derivation of the Tomita relation for no-flux boundary conditions}
\label{app:noflux}
In this Appendix, we show how the relationship between $\A$ and $\rcom^2$ is derived in detail for a system with no-flux boundary conditions, paying attention to the limits of integration. For no-flux boundary conditions, the self-correlation function is defined as:
\begin{align}
C(\xv) &= \frac{1}{L^{d}}\int_{\mathrm{max}(-\xv,0)}^{\mathrm{min}(L-\xv,L)}\df(\xv+\yv)\df(\yv)\dd[d]{\bf y},
\end{align}
where the components of $\xv$ can take values between $-L$ and $L$ and the limits of the $\bf y$ integral, to be thought of component-wise, are chosen to ensure that $\df(\xv+\yv)$ is well-defined and that the entire system of volume $L^d$ is integrated over no matter the sign of the components in $\xv$.

The structure factor is then defined as:
\begin{equation}
    S(\qv) = \int_{-L}^{L}\dd[d]{\xv}C(\xv)e^{i\qv\vdot\xv},
\end{equation}
and the coefficient of $q^2$ reads:
\begin{align}
\A &= -\frac{1}{2dL^{d}}\int_{-L}^L\dd[d]{\xv}\int_{\mathrm{max}(-\xv,0)}^{\mathrm{min}(L-\xv,L)}\dd[d]{\bf y}\df(\xv+\yv)\df(\yv)\xv^2.
\end{align}
It can be rewritten in terms of $\rcom$ as follows. First, we split up the $x_i$ integrals into their positive and negative parts, and transform $x_i\to -x_i$ in the latter parts:
\begin{align}
\A &= -\frac{1}{2dL^{d}}\prod\limits_{i=1}^d\left[\int_0^L\dd{x_i}\int_0^{L-x_i}\dd{y_i}+\int_{-L}^0\dd{x_i}\int_{-x_i}^{L}\dd{y_i}\right]\xv^2\df(\vb{x}+\yv)\df(\yv)\nonumber\\
&= -\frac{1}{2dL^{d}}\int_0^L\dd[d]{\xv}\xv^2\nonumber\\
&\phantom{{}={}-}\prod\limits_{i=1}^d\left[\int_0^{L-x_i}\dd{y_i}\int^L_0\dd{z_i}\delta(z_i-x_i)+\int_{x_i}^{L}\dd{y_i}\int^L_0\dd{z_i}\delta(z_i+x_i)\right]\df(\vb{z}+\yv)\df(\yv).
\end{align}
Now we switch the domains of integration of $\xv$ and $\yv$ and transform $x_i\to x_i \pm y_i$.
\begin{align}
\A
&= -\frac{1}{2dL^{d}}\int_0^L\dd[d]{\bf y}\nonumber\\
&\phantom{{}={}-}\prod\limits_{i=1}^d\left[\int_0^{L-y_i}\dd{x_i}\int^L_0\dd{z_i}\delta(z_i-x_i)+\int_{0}^{y_i}\dd{x_i}\int^L_0\dd{z_i}\delta(z_i+x_i)\right]\xv^2\df(\vb{z}+\yv)\df(\yv)\nonumber\\
&= -\frac{1}{2dL^{d}}\int_0^L\dd[d]{\bf y}\prod\limits_{i=1}^d\left[\int_{y_i}^{L}\dd{x_i}+\int_{0}^{y_i}\dd{x_i}\right](\xv-\yv)^2\df(\xv)\df(\yv).
\end{align}
The two integrals over $x_i$ can now be joined together. Finally, using that the integral of $\df$ over the entire system vanishes, we obtain the desired relation:
\begin{align}
\A
&= -\frac{1}{2dL^{d}}\int_0^L\dd[d]{\xv}\int_0^L\dd[d]{\bf y}(\xv-\yv)^2\df(\xv)\df(\yv)\nonumber\\
&= \frac{1}{dL^{d}}\int_0^L\dd[d]{\xv}\df(\xv)\xv\vdot\int_0^L\dd[d]{\bf y}\yv\df(\yv) = \frac{L^d}{d}\rcom^2\,.
\end{align}

\section{Derivation of the $S_0$ contribution to the centre of mass vector}
\label{app:S0}
In this Appendix, we show how the relation \eqref{eq:A_R} is modified when $S_0\neq 0$, i.e. for non-hyperuniform initial conditions, giving rise to \eqref{eq:A_R_S0}.
We work in Fourier space, using the convention:
\begin{subequations}
\begin{align}
    \df_\qv &= \frac{1}{L^d}\int\dd[d]{\xv}\df(\xv)e^{i\qv\vdot\xv},\\
    \df(\xv) &= \sum_{\qv\in \frac{2\pi}{L}\mathbb{Z}^d}\df_\qv e^{-i\qv\vdot\xv}.
\end{align}
\end{subequations}
Note that $\df_{\qv=0}=0$ by definition.

Using the identity derived in \ref{app:noflux}, the centre of mass of the field $\df(\xv)$ can be written as:
\begin{equation}
\rcom^2 = -\frac{1}{2L^{2d}}\int\dd[d]{\xv}\dd[d]{\vb{y}}\df(\xv+\yv)\df(\yv)\xv^2,
\end{equation}
Inserting the Fourier expansion of $\df$ and evaluating the $\yv$ integral, this gives:
\begin{equation}
\rcom^2 = -\frac{1}{2L^{d}}\int\dd[d]{\xv}\sum_{\qv}\abs{\df_\qv}^2 e^{-i\qv\vdot\xv}\xv^2.
\end{equation}
Now, assuming isotropy, in Fourier space the structure factor \eqref{eq:S} reads $S(q) = L^d\abs{\df_\qv}^2$. The zero mean of $\df(\xv)$ implies $S(q=0) = 0$. On the other hand, for positive $q>0$, we Taylor expand the structure factor as in \eqref{eq:pre-quench}, $S(q>0) = S_0 + S_2q^2 + S_4q^4$, which gives:
\begin{equation}
\rcom^2 = -\frac{1}{2L^{2d}}\int\dd[d]{\xv}\sum_{\qv\neq \vb{0}}(S_0 + S_2q^2 + S_4q^4) e^{-i\qv\vdot\xv}\xv^2.
\end{equation}
Next, we use the identities $\sum_{\qv\neq \vb{0}}e^{-i\qv\vdot\xv} = L^d\delta(\xv) - 1$ and $\sum_{\qv\neq \vb{0}}\qv^ne^{-i\qv\vdot\xv} = L^d(i\grad)^n\delta(\xv)$ to write:
\begin{equation}
\rcom^2 = -\frac{1}{2L^{d}}\int\dd[d]{\xv}[S_0(\delta(\xv)-L^{-d}) - S_2\laplacian\delta(\xv) + S_4\nabla^4\delta(\xv)]\xv^2.
\end{equation}
Using $S_0\delta(\xv)\xv^2=0$ and integrating by parts, we obtain:
\begin{equation}
\rcom^2 = \frac{1}{2L^{d}}\int\dd[d]{\xv}[S_0L^{-d}\xv^2 + S_2\delta(\xv)\laplacian\xv^2],
\end{equation}
which, with $\laplacian\xv^2=2d$, finally gives:
\begin{equation}
\label{eq:A}
\rcom^2 \sim dL^{-d-1}L^3 S_0 + dL^{-d}S_2,
\end{equation}
so that \eqref{eq:A_R_S0} holds.

\section{Structure factor in the Gaussian C-DP model}
\label{app:C-DP}
In this Appendix, we show how to obtain the structure factor solution \eqref{eq:C-DP_S} from \eqref{eq:C-DP}. To this goal, we consider the dynamics of the fields in Fourier space. Using $\df = \rho - (a_0+p_0)$, and $\df^A = \rho_A - a_0$, the transformed dynamics read:
\begin{equation}
\label{eq:C_DP_app1}
    \pdv{t}\mqty(\df^A_\qv\\\df_\qv) = \underbrace{\mqty(-\alpha q^2-r& r\\ -\alpha q^2 & 0)}_{:=\,\vb{M}}\mqty(\df^A_\qv\\\df_\qv) + \mqty(\sqrt{2T}\tilde\omega_\qv-i\qv\vdot\Lv_\qv\\-i\qv\vdot\Lv_\qv).
\end{equation}
Diagonalising the matrix $\vb{M}$ appearing in the dynamics gives the eigenvectors $(1,1)$ with eigenvalue $-\alpha q^2$ and $(r, \alpha q^2)$ with eigenvalue $-r$. The solution of \eqref{eq:C_DP_app1} is then given by:
\begin{align}
\label{eq:C_DP_app2}
    \mqty(\df^A_\qv(t)\\\df_\qv(t)) &= \tilde{c}_1(\qv)e^{-\alpha q^2t}\mqty(1\\1) + \tilde{c}_2(\qv)e^{-rt}\mqty(r\\\alpha q^2)\nonumber\\
    &\phantom{{}={}}+ e^{\vb{M}t}\int_{0}^t\dd{t^\prime}e^{-\vb{M}t^\prime}\mqty(\sqrt{2T}\tilde\omega_\qv(t^\prime)-i\qv\vdot\Lv_\qv(t^\prime)\\-i\qv\vdot\Lv_\qv(t^\prime)),
\end{align}
where $\tilde{c}_1(\qv)$ and $\tilde{c}_2(\qv)$ are determined by the initial conditions. Assuming these are isotropic, $\tilde{c}_{1,2}(\qv) = \tilde{c}_{1,2}(q)$. We decompose the noise vector in the second line into eigenvectors of $\vb{M}$,
\begin{align}
    \mqty(\sqrt{2T}\tilde\omega_\qv-i\qv\vdot\Lv_\qv\\-i\qv\vdot\Lv_\qv(t^\prime))(t) &= b_1(\qv,t)\mqty(1\\1) + b_2(\qv,t)\mqty(r\\\alpha q^2),
\end{align}
where the coefficients $b_{1,2}(\qv,t)$ read:
\begin{subequations}
\begin{align}
    b_1(\qv,t) &= -\frac{\sqrt{2T}}{r-\alpha q^2}\alpha q^2\tilde\omega_\qv -i\qv\vdot\Lv_\qv,\\
    b_2(\qv,t) &= \frac{\sqrt{2T}}{r-\alpha q^2}\tilde\omega_\qv.
\end{align}
\end{subequations}
The decomposition simplifies the exponentials in \eqref{eq:C_DP_app2}, so that the Fourier components of the total density obey:
\begin{equation}
    \df_\qv(t) = \tilde{c}_1(q)e^{-\alpha q^2t} + \alpha q^2\tilde{c}_2(q)e^{-rt} + \int_{0}^t\dd{t^\prime}[e^{\alpha q^2(t^\prime-t)}b_1(\qv,t^\prime) + \alpha q^2e^{r(t^\prime-t)}b_2(\qv,t^\prime)].
\end{equation}
The structure factor $S(\qv) = L^d\expval{\df_\qv\df_{-\qv}}$ is then:
\begin{align}
    L^{-d}S(\qv) &= [\tilde{c}_1(q)e^{-\alpha q^2t} + \alpha q^2\tilde{c}_2(q)e^{-rt}]^2\nonumber\\
    &\phantom{{}={}}+ \int_{0}^t\int_{0}^t\dd{t^\prime}\dd{t^{\prime\prime}}[e^{\alpha q^2(t^{\prime\prime} + t^\prime-2t)}\expval{b_1(\qv,t^\prime)b_1(-\qv,t^{\prime\prime})}\nonumber\\
    &\phantom{{}={}}\hspace{2em}+ \alpha^2 q^4e^{r(t^{\prime\prime} + t^\prime-2t)}\expval{b_2(\qv,t^\prime)b_2(-\qv,t^{\prime\prime})}\nonumber\\
    &\phantom{{}={}}\hspace{2em}+ 2\alpha q^2e^{\alpha q^2(t^\prime-t)+r(t^{\prime\prime} -t)}\expval{b_1(\qv,t^\prime)b_2(-\qv,t^{\prime\prime})}]. \label{eq:C_DP_app3}
\end{align}
Using the noise variances in Fourier space,
\begin{align}
    \expval{\tilde{\omega}(\qv,t)\tilde{\omega}(\kv,t^{\prime})} &= \frac{1}{L^d}\delta_{\qv,-\kv}\delta(t-t^\prime),\\
    \expval{\Lambda_i(\qv,t)\Lambda_j(\kv,t^{\prime})} &= \frac{2D}{L^d}\delta_{ij}\delta_{\qv,-\kv}\delta(t-t^\prime),
\end{align}
we find
\begin{align}
    \expval{b_1(\qv,t^\prime)b_1(-\qv,t^{\prime\prime})} &= \frac{1}{L^d}\delta(t^\prime-t^{\prime\prime})\left[\frac{2T}{(r-\alpha q^2)^2}\alpha^2q^4+2D q^2\right],\\
    \expval{b_2(\qv,t^\prime)b_2(-\qv,t^{\prime\prime})} &= \frac{1}{L^d}\delta(t^\prime-t^{\prime\prime})\left[\frac{2T}{(r-\alpha q^2)^2}\right],\\
    \expval{b_1(\qv,t^\prime)b_2(-\qv,t^{\prime\prime})} &= \frac{1}{L^d}\delta(t^\prime-t^{\prime\prime})\left[-\frac{2T\alpha q^2}{(r-\alpha q^2)^2}\right].
\end{align}
Inserting these expressions into \eqref{eq:C_DP_app3}, and defining $c_{1,2}=L^{d/2}\tilde{c}_{1,2}$, we obtain the structure factor \eqref{eq:C-DP_S} given in the main text.

We now show that the intermediate-time scaling discussed in the main text maintains the same qualitative features when time scale separation is present. If, instead of $r \sim \alpha L^{-2}$, we are in the fast reaction regime, and consider times $r^{-1}\ll t \ll (\alpha q^2)^{-1}$, the expression \eqref{eq:C-DP_S} reduces to:
\begin{align}
    S(q) &= c_1(q)^2 + 2Dq^2t - 2c_1(q)^2\alpha q^2t + \frac{2T\alpha^2 q^4}{(r-\alpha q^2)^2}t \nonumber\\
    &+ \frac{\alpha^2q^4T}{(r-\alpha q^2)^2}\left(\frac{1}{r}-\frac{4}{r+\alpha q^2}\right) + O((\alpha q^2t)^2).
\end{align}
Here in the fast reaction regime, the reaction term has settled into a time-independent contribution, with a leading $q^4$-order. On the other hand, in the slow reaction limit,  $t\ll (\alpha q^2)^{-1}\ll r^{-1}$, and:
\begin{align}
    S(q,t) &= S(q,0) + 2D q^2t - 2\sqrt{S(q,0)}c_1(q)\alpha q^2t + O((\alpha q^2t)^2,rt)
\end{align}
Hence, the structure factor has the same scaling for $t\ll (\alpha q^2)^{-1}$ no matter the choice of $r$: a non-vanishing standard noise $D > 0$ gives $\qexp = 2$ at intermediate times, whereas the initial quartic scaling, $\qexp = 4$, survives on these time scales for $D = 0$.
\pagebreak

\end{document}